\documentclass[fleqn]{article}
\usepackage[a4paper,margin=25.4mm]{geometry}
\usepackage[english]{babel}

\usepackage{amsmath}
\usepackage{amssymb}
\usepackage{sistyle}
\usepackage{cite}

\usepackage{graphicx}			
\usepackage{wrapfig}			
\usepackage[
	labelfont={sf,bf},			
	textfont={sf},			
	font=small,					
	justification=RaggedRight	
	]{caption}
\usepackage{subcaption}			
\usepackage{epstopdf}			
\usepackage{epsfig}
\usepackage{url}
\usepackage{listings}
\usepackage{xcolor}

\usepackage[unicode=true,pdfusetitle, bookmarks=false, breaklinks=false,pdfborder={0 0 0},backref=false,colorlinks=false] {hyperref}

\newcounter{codesnip}[section]

\begin{document}

\begin{center}
	\hfill\\[1.0cm]
	{\Large\textbf{Quantum detector tomography applied to the human visual system: a feasibility study}}\\[0.2cm]
	{\large{T.H.A. van der Reep$\left.^{1,2,*}\right.$, D. Molenaar$\left.^{1}\right.$, W. L\"offler$\left.^{2}\right.$ and Y. Pinto$\left.^{1}\right.$}} \\[0.2cm]
	{$\left.^{1}\right.$\emph{Psychology Research Institute, Nieuwe Achtergracht $129$b, $1018$ WS Amsterdam, The Netherlands}}\\[0.1cm]
	{$\left.^{2}\right.$\emph{Leiden Institute of Physics, Niels Bohrweg $2$, $2333$ CA Leiden, The Netherlands}}\\[0.1cm]
	$\left.^{*}\right.$t.h.a.vanderreep@uva.nl\\[0.2cm]
	\today
\end{center}
\begin{abstract}
We show that quantum detector tomography can be applied to the human visual system to explore human perception of photon number states. In detector tomography, instead of using very hard to produce photon number states, the response of a detector to light pulses with known photon statistics of varying intensity is recorded, and a model is fitted to the experimental outcomes thereby inferring the detector's photon number state response. Generally, light pulses containing a Poisson-distributed number of photons are utilised, which are very easy to produce in the lab. This technique has not been explored to study the human visual system before, because it usually requires a very large number of repetitions not suitable for experiments on humans. Yet, in the present study we show that detector tomography is feasible for human experiments. Assuming a simple model for this accuracy, the results of our simulations show that detector tomography is able to reconstruct the model using Bayesian inference with as little as $5000$ trials. We then optimize the experimental parameters in order to maximise the probability of showing that the single-photon accuracy is above chance. As such, our study opens the road to study human perception on quantum level.
\end{abstract}

\section{Introduction}\label{secIntroduction}
With the advancement of quantum optics, detectors have been developed that are sensitive to single photons. However, determining \emph{how} sensitive the detectors are to $n$-photon number states is a non-trivial problem, since a light source producing photon number states is not available. This difficulty has been overcome by quantum detector tomography techniques \cite{Lundeenetal2009,Feitoetal2009,Renemaetal2012}, in which the detector's ``clicking'' probability is inferred by irradiating it with light pulses which are readily available, such as pulses with Poissonian photon statistics from a laser.\\
A similar case can be made for the detection of light by the human visual system: How well can humans perceive few-photon states? The challenge for detecting such states is to overcome the intrinsic noise in the visual system. To our knowledge, this is the first time a quantum technique such as detector tomography is considered for studying human perception, although it has been proposed to use quantum biometry as a secure identification process \cite{Loulakis2017}. Since the $1940$s it is known from experiments by Hecht \emph{et al.} that the human visual system is sensitive to light pulses containing only a few photons \cite{Hechtetal1941}. In later experiments similar and lower limits have been found, down to single-photon level \cite{Sakitt1972, Teichetal1982}. Recently, Tinsley \emph {et al.} presented evidence that humans are indeed able to detect single photons with an accuracy above chance \cite{Tinsleyetal2016}. This is quite remarkable, given that the overall efficiency of the human eye, from cornea up to producing a retinal signal, is only $0.1-0.4$ \cite{Barlow1977,Hallett1987,Donner1992} and a single photon will trigger a single rhodopsin molecule only \cite{Loulakis2017}, which has to be amplified and read out in the noisy environment of the brain. Knowing the limits of human visual perception informs us about the boundaries of the perceptual machinery. This, in turn, is fundamental for our understanding of how the brain generates conscious perception. For instance, to enable conscious awareness of just one photon, specific signal-to-noise mechanisms (e.g. without broad averaging) seem required.\\ 
In their analysis, Hecht \emph{et al.} assumed a step function: $n$-photon states are either imperceivable ($n<n_{\text{crit}}$) or fully perceivable ($n\geq n_{\text{crit}}$), whereas Tinsley \emph{et al.} only considered the perceptibility of single photons. In the present study we will consider detector tomography as a solution to bridge the gap in these two approaches and determine the perceptibility of few-photon states. Apart from this, we can use the techniques to optimise our experimental design, i.e., using our proposed technique, we determine the experimental parameters necessary to statistically demonstrate above-chance performance in a single photon perception task. We implement our technique in a Bayesian modeling framework which has a number of advantages over a frequentist approach (see e.g., \cite{Wagenmakersetal2008}). For the present study, the most important reason to rely on Bayesian statistics is that it allows quantification of all relevant hypotheses (i.e., the null-hypothesis of chance performance and the alternative hypothesis of above chance performance). Using frequentist significance tests, it is only possible to quantify evidence against the null-hypothesis such that the null-hypothesis can never be confirmed. Note that despite this advantage of Bayesian statistics over frequentist statistics, a Bayesian approach to hypothesis testing has not been applied to detector tomography yet.\\

Our setting is the following: we consider a single test subject that performs a $2$-alternative forced choice ($2$AFC) task (see section \ref{sec2AFC}). During the trials, Poisson-distributed few-photon states are sent to the subject's eye in one of two possible intervals. The subject has to indicate which interval the pulse was sent from which the accuracy can be determined. We assume that the pulses have a wavelength of $\SI{500}{nm}$, at which the quantum efficiency of the rods in the retina is maximal. The pulses are sent towards the location on the retina where rods are most abundant ($16^{\circ}-23^{\circ}$ away from the yellow spot \cite{Cursioetal1990,Tinsleyetal2016,Holmesetal2017}).\\
The difficulty of this setting lies in the fact that the repetition rate of the experiment is low. Where detector tomography for, e.g., avalanche photo-diodes or superconducting photo-detectors can be collected at a rate in the range $\SI{}{kHz}-\SI{}{MHz}$ respectively \cite{Lundeenetal2009,Renemaetal2012}, typical repetition rates for human test subjects are $0.1-\SI{1}{Hz}$ \cite{Tinsleyetal2016,Holmesetal2017}. Human subjects have a limited attention span which further decreases the data collection rate (data collection sessions can last for presumably $\sim\SI{2}{h}$ maximally). This implies we would like to measure at low average photon number only, whereas ordinarily detector tomography is applied to measurements at average photon numbers ranging from imperceivable to always perceivable.\\

In this work we show that detector tomography is feasible with a human test subject by simulating the proposed experiment. Apart from that, we determine the optimal experimental parameters for excluding imperceptibility of single photons, given that single photons can be perceived by (modelled) humans with an accuracy above chance level. First, we explain $2$AFC tasks and detector tomography with Poissonian light pulses. In section \ref{secEyeModel} we develop a simple visual perception model describing the $n$-photon accuracies, since to our knowledge such a model has not been considered yet. Section \ref{secSimulationReconstruction} describes the computer simulation, after which we dive into reconstruction (parameter recovery) of the model from our simulation data. Here we apply detector tomography in the framework of Bayesian inference. In section \ref{secSingleSim} we show that the reconstruction returns $n$-photon accuracy values in agreement with our model. We continue by determining the optimal experimental parameters for excluding imperceptibility of single photons in section \ref{secOptimise}. In this section we also determine the influence of our visual detection model and the influence of noise on the average photon number emitted by the light source. We discuss our results and extensions to our simulation and reconstruction programs in section \ref{secDiscussion}, after which we conclude in section \ref{secConclusion}.

\section{Two-alternative forced choice tasks}\label{sec2AFC}
\begin{figure}[htbp]
	\centering
	\begin{subfigure}[b]{0.30\textwidth}
		\includegraphics[width=\textwidth]{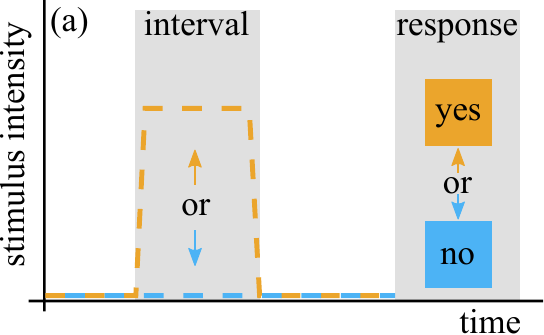}
		\label{sfig_YN}
	\end{subfigure}\hspace{10mm}
	\begin{subfigure}[b]{0.40\textwidth}
		\includegraphics[width=\textwidth]{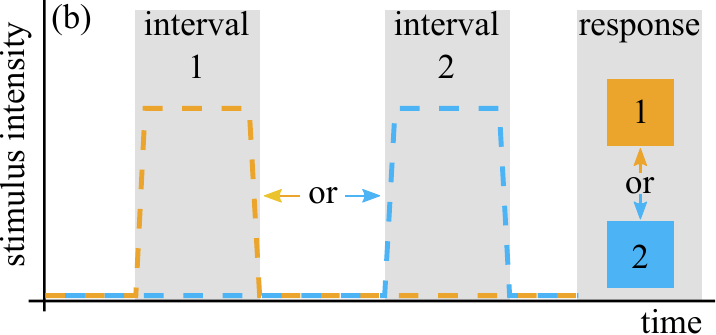}
		\label{sfig_2AFC}
	\end{subfigure}\vspace{-5mm}
	\caption{Schematic overview of (a) a yes/no task and (b) a (time separated) $2$AFC task. In a yes/no trial a stimulus is either present or absent in a single interval and the subject indicates whether a stimulus is detected. In a $2$AFC trial a stimulus is presented in one of two intervals and a test subject indicates in which interval the stimulus was presented.}\label{fig2AFC_vs_YN}
\end{figure}

In psychophysics, the quantitative study of relations between physical and psychological events, $2$AFC tasks are often employed as a technique to determine the limits of perception \cite{Fechner1860,KingdomPrins2016}. Intuitively, the probability $P$ of perceiving a weak light pulse can be determined in a series of trials, in which randomly the pulse is presented or not, see figure \ref{fig2AFC_vs_YN}a. For each trial subjects are asked whether a pulse was present or not, which is referred to as a yes/no task in psychophysics.\\ 
This contrasts with the $2$AFC task measuring a subject's accuracy $A$ of perceiving the pulse of light. In every trial of such tasks subjects are presented with a pulse in one of two intervals, whereas in the other interval no pulse is presented, see figure \ref{fig2AFC_vs_YN}b. The interval in which the stimulus is presented is chosen randomly, each with probability $1/2$. These intervals can be separated in space (did the stimulus come from left or right?) or in time (was the stimulus presented early or late? This is also referred to as a two-interval forced choice ($2$IFC) task). After a $2$AFC trial, the subjects are forced to select the interval in which they believe the actual light pulse was presented.\\

$P$ and $A$ are -- in theory -- related as
\begin{equation}\label{eqPtoA}
	A_{\text{}}=\dfrac{1}{2}\left(1+P_{\text{}}\right).
\end{equation}
The lower limit of $A_{\text{}}$ of $1/2$ arises, because if subjects do not detect the stimulus, a random guess has to be made in which interval the stimulus was presented. An accuracy in excess of $1/2$ implies the test subject is able to detect the stimulus with a non-zero probability. Finally, if subjects always detects a stimulus, $A_{\text{}}$ and $P_{\text{}}$ both equal $1$.\\
In practice, however, equation (\ref{eqPtoA}) does not hold necessarily. Yes/no tasks are more susceptible to response bias. The decision threshold to answer ``yes'' in such experiments is left free to the subjects, who may choose to rate the various trials anywhere in the range from strict to loose. Subjects may even alter their decision threshold (un)willingly, which greatly influences the experimental data, see e.g.\cite{Teichetal1982}. Alternatively, in $2$AFC tasks subjects rate the difference in sensory input between the two intervals, tackling this issue. Although $2$AFC experiments are influenced by interval bias, the influence of this bias can be minimised \cite{YeshurunCarrascoMaloney2008,GarciaAlcala2011}. For this reason we will consider a $2$AFC task in this study.

\section{Detector tomography with Poissonian light pulses}\label{secDetectorTomography}
Detector tomography can be used to obtain the $n$-photon accuracy of a test subject for a range of $n$-photon states, while not possessing a source producing these states deterministically. Instead, one uses a source for which the photon-number distribution is known and uses statistics to infer the $n$-photon accuracies. Hence, let us consider a light source with a known photon-number distribution. I.e., we know the probability that the source presents $n=0,1,2,\dots$ photons to the subject. In such a case, the subject's accuracy for some constant source setting with known photon-number distribution is given by \cite{Renemaetal2012}\footnote{Note that a similar equation would arise for a yes/no task in an experiment measuring detection probability, in which the subject merely notes whether a light pulse was observed.}
\begin{equation}\label{eqAccuracyGen}
\begin{aligned}
	A\left(I_{\text{s}}\right)&=\sum_{n=0}^\infty a_n\rho_n\left(I_{\text{s}}\right)\\
	&=1-\sum_{n=0}^\infty \left(1-a_n\right)\rho_n\left(I_{\text{s}}\right)\\
	&\approx 1-\sum_{n=0}^{n_{\text{max}}} \left(1-a_n\right)\rho_n\left(I_{\text{s}}\right).
\end{aligned}
\end{equation}
In this equation, $I_{\text{s}}$ are the parameters that determine the photon number distribution of the source, $a_n$ is the accuracy of the subject for \textit{exactly} $n$ photons and $\rho_n\left(I_{\text{s}}\right)$ is the probability that the source with settings $I_{\text{s}}$ emits $n$ photons. We rewrite this equation in the second line straightforwardly to terminate the sum in practice: When a stimulus with $n_{\text{max}}+1$ photons is always detected ($a_{n_{\text{max}}+1}=1$) or when the experiment is designed such that trials with $n_{\text{max}}+1$ photons are practically not present, whereas trials with $n=0,\dots,n_{\text{max}}$ do occur ($\rho_{0,\dots, n_{\text{max}}}> 0$, $\rho_{n_{\text{max}}+1}\approx 0$), the sum terminates. This is stipulated in the third line of the equation, that introduces the parameter $n_{\text{max}}$ explicitly.\\ 
For Poissonian light sources the photon number distribution is determined by
\begin{equation}\label{eqPoisson}
	\rho_n=\exp\left(-\bar{N}\right)\dfrac{\bar{N}^n}{n!}.
\end{equation}
Here, $\bar{N}$ is the average number of photons per light pulse. Substituting equation (\ref{eqPoisson}) into equation (\ref{eqAccuracyGen}) we arrive at an accuracy for mean photon number $\bar{N}$ of
\begin{equation}\label{eqAccuracyP}
	A\left(\bar{N}\right)=1-\exp\left(-\bar{N}\right)\sum_{n=0}^{n_{\text{max}}} \left(1-a_n\right)\dfrac{\bar{N}^n}{n!}.
\end{equation}

In a $2$AFC task $A\left(I_{\text{s}}\right)$ is measured and $\rho_n\left(I_{\text{s}}\right)$ is known for several $I_{\text{s}}$. This allows to reconstruct $\vec{a}=\left[a_0,a_1,\dots, a_{n_{\text{max}}}\right]$ by fitting equation (\ref{eqAccuracyGen}) to the measured data. For light pulses with Poissonian photon statistics one would fit equation (\ref{eqAccuracyP}), of course. Finding $\vec{a}$ and $n_{\text{max}}$ is the goal of detector tomography, a process which is further described in section \ref{secSimulationReconstruction}.

\section{Visual perception model}\label{secEyeModel}
The model describing the subject's accuracy (for constant source setting) $A$, given the source's photon number distribution $\rho_n$ presented in equation (\ref{eqAccuracyGen}), depends heavily on the $n$-photon accuracies $a_n$. Of course, $a_0=0.5$ due to a lack of photons being presented to the subject and \cite{Tinsleyetal2016} estimates $a_1\approx 0.516$. To our knowledge, however, no experimental bounds have been obtained for $a_{>1}$. Thus, in order to perform the feasibility study, which is the main topic of this study, we need to construct a model for $\vec{a}$: the visual perception model.\\

In order to construct the model, let us consider a Gaussian pulse of $n$ photons with a wavelength of $\SI{500}{nm}$. If such a pulse is focussed on the retina, these photons land on the retina within an area of approximately $S=\pi w_0^2$, where the beam waist
\begin{equation}
	w_0=\dfrac{\lambda}{\pi n_{\text{eye}}\theta}\approx\SI{3}{\mu m}.
\end{equation}
In this equation, $n_{\text{eye}}\approx 1.337$ is the refractive index of the eye and $\theta$ is the convergence angle of the pulse. If the pulse of collimated light is focussed by a lens with a $\SI{2}{cm}$-radius and a focus distance of $\SI{50}{cm}$, $\theta\approx \SI{4e-2}{rad}$. This implies $S\approx\SI{3e1}{\mu m^2}$. From \cite{Cursioetal1990} we estimate that a single rod cell covers approximately $\SI{5}{\mu m^2}$ of retinal area, implying the pulse covers approximately $6$ rod cells.\\
If on the other hand a Maxwellian view \cite{Westheimer1966} is used during the experiments, in which the pulse is focussed on the eye lens (instead of the retina) and a small area of the retina is irradiated, $S\approx\pi (d_{\text{eye}}\theta/n_{\text{eye}})^2\approx\SI{1.(6)e6}{\mu m^2}$, assuming the diametre of the eye equals $\SI{24}{mm}$. In this case the pulse covers approximately $\SI{3e5}{}$ rod cells.\\

We intend to use the Maxwellian view during our experiments, which maximises the number of photons from the pulse entering the eye. Additionally, focussing the pulse on the eye lens prevents this lens from influencing the path of the photons, such that we do not need to take precautions fixating the focal distance of the eye. Given that we will consider approximately $10$ photons per pulse at most, the calculation presented before implies that in this case it is likely that all photons reach a different rod. If we furthermore assume that at such light levels all rods function independently, we are led to a binomial model for $a_n$. I.e. we set a single photon detection probability, $p_1$, and calculate
\begin{equation}\label{eqBinomModel}
	p_n=1-\left(1-p_1\right)^n.
\end{equation}
From this equation, 
\begin{equation}\label{eqptoa}
	a_n=\dfrac{1}{2}\left(1+p_n\right)
\end{equation}
similar to equation (\ref{eqPtoA}), which is the final necessity to simulate an experiment. We note that the reconstruction method that will be discussed in section \ref{secSimulationReconstruction} does not depend on the model we constructed for $a_n$.

\section{Simulation and reconstruction}\label{secSimulationReconstruction}
To simulate the envisioned experiment, we have written a simulation program in \textsc{R} \cite{Rcore2017}. In this program, we set the minimum and maximum intensity $\bar{N}_{\text{min}}$ and $\bar{N}_{\text{max}}$, which range is divided in $D$ equidistant data points. Together these are the $\vec{\bar{N}}$ intensities for the Poissonian light source. Additionally, we set the number of trials per data point $d$, $T$ and the noise of the light source. We consider the following noise model
\begin{equation}
	\bar{N}_d=\bar{N}_{d,0}+\mathrm{d}\bar{N}_d,
\end{equation}
where $\mathrm{d}\bar{N}_d\sim\mathcal{N}(0,\sigma_{\bar{N},d})$. I.e., for all trials, we add a normally distributed random deviation ($0$-mean, $\sigma_{\bar{N},d}$-standard deviation) to the intended mean photon number $\bar{N}_{d,0}$. For each of the data points, we calculate the theoretical accuracy $A$ by setting the single-photon detection probability $p_1$ and using equations (\ref{eqBinomModel}), (\ref{eqptoa}) and (\ref{eqAccuracyP}).\\
As a next step, we simulate the trials. For each trial for each data point $d$ (at source intensity $\bar{N}_{d,0}$), we draw a source intensity deviation $\mathrm{d}\bar{N}_d$. Sequentially, we draw $n$ photons from a Poisson distribution with mean photon number $\bar{N}_{d}$. $n$ represents the number of photons presented to the subject's eye. Assuming the subject to be unbiased and performing Bernoulli trials, we draw the trial outcome from a Bernoulli distribution with accuracy $a_n$, see equation (\ref{eqptoa}). This results in either a correct or a wrong response. We sum the number of correct responses per data point and thus obtain a $1$-dimensional array of length $D$, which elements are the number of correct responses for data point $d$ at mean photon number $\bar{N}_{d,0}$. This array we will refer to as $\vec{\Sigma}$.\\

Using the data array with trial outcomes, our goal is to obtain $n_{\text{max}}$ and reconstruct the model accuracies $\vec{a}=\left[a_0,a_1,\dots,a_{n_{\text{max}}}\right]$ set by the user. Here we give a summary of the methods. The interested reader may refer to appendix \ref{secReconstruction} for a more elaborate discussion.\\
For a given $n_{\text{max}}$, we perform the reconstruction of $\vec{a}$ under Bayesian inference using a program written in \textsc{RStan} \cite{Rstan}. The program takes $\vec{\Sigma}$ as input. Apart from this array, the number of data points $D$, the number of trials per data point $T$, $n_{\text{max}}$ and the $\boldsymbol{\rho}$-matrix need to be specified. The latter is a $D\times (n_{\text{max}}+1)$-matrix whose indices are given by $\boldsymbol{\rho}_{d,n}=\exp(-\bar{N}_d)\bar{N}_d^n/n!$, i.e., $\boldsymbol{\rho}$ contains the theoretical probabilities for presenting $n=0,\dots,n_{\text{max}}$ photons.\\
Using this program, we obtain parameter estimates for $\vec{a}$ by Markov Chain Monte-Carlo (MCMC) sampling from the posterior distribution of these unknown parameters,
\begin{equation}\label{eqBayesRule}
	p\left(\vec{\tilde{a}}|\vec{\Sigma}\right)\propto p\left(\vec{\Sigma}|T,\vec{\tilde{a}}\right)p\left(\vec{\tilde{a}}\right),
\end{equation}
and taking the posterior mode of these samples. The resulting Markov Chains (MCs) of samples from the posterior distribution have a length
\begin{equation}\label{eqNmc}
	N_{\text{MC}}=\dfrac{N_{\text{chains}}\left(N_{\text{iter}}-N_{\text{warmup}}\right)}{N_{\text{thin}}}.
\end{equation}
where $N_{\text{chains}}$ is the amount of parallel chains evaluated (in this study typically $3$), $N_{\text{iter}}$ is the number of iterations (in this study typically $15000$), $N_{\text{warmup}}$ the number of initial iterations discarded to ensure convergence of the sampling algorithm (in this study typically $2500$) and $N_{\text{thin}}$ the amount of thinning in the chain decrease the autocorrelations among the samples (in this study typically $3$).\\
In equation (\ref{eqBayesRule}) $\vec{\tilde{a}}$ is the reconstruction of $\vec{a}$. $p(\vec{\tilde{a}})$ is the prior distribution of $\vec{\tilde{a}}$ (which we will denote $p(\vec{\tilde{a}}^{(0)})$ from now on) and $p(\vec{\Sigma}|T,\vec{\tilde{a}})$ is the likelihood of the data $\vec{\Sigma}$ given $T$ and $\vec{\tilde{a}}$. For the posterior $p(\vec{\tilde{a}}|\vec{\Sigma})$ we will use the shorthand notation $p(\vec{\tilde{a}}^{(1)})$. Below we discuss the prior distribution and likelihood in more detail. This technical description is added here for completeness, but readers may skip ahead to section \ref{secSingleSim}.\\ 

For $p(\vec{\tilde{a}}^{(0)})$ we choose a prior distribution determined by a $\text{beta}(\alpha,\beta)$-distribution with hyper-prior shape parameters
\begin{align}\label{eqBetaScalePriors}
	\alpha&\sim\mathrm{halfnorm}\left(1,1/2+n_{\text{max}}/2-n_{\text{max}}^2/35\right),\\
	\beta&\sim\mathrm{halfnorm}\left(1,12-n_{\text{max}}/1.8-n_{\text{max}}^2/200\right).
\end{align}
As discussed in appendix \ref{ssecPrior}, the $\text{beta}(\alpha,\beta)$-distribution for specific $\alpha$ and $\beta$ can be seen as a prior model for the detection probability $\vec{\tilde{p}}=2\vec{\tilde{a}}-1$ (equation (\ref{eqptoa})). By drawing sets of $n_{\text{max}}$ samples from the prior beta-distribution, where each set is sorted and appended to $0$ (thus yielding the drawn prior vector $\vec{\tilde{p}}^{(0)}=[0,\tilde{p}_1^{(0)},\dots,\tilde{p}_{n_{\text{max}}}^{(0)}]$ ($\tilde{p}_n^{(0)}<\tilde{p}_{n+1}^{(0)}$ due to sorting)) and finally transformed as $\vec{\tilde{a}}^{(0)}=(1+\vec{\tilde{p}}^{(0)})/2$, we find that the distribution $p(\vec{\tilde{a}}^{(0)}$) complies with our common sense and current knowledge of $\vec{a}$ (see appendix \ref{ssecPrior}): $\tilde{a}_0^{(0)}=1/2$ ($0$ photons are imperceivable), $\tilde{a}_n^{(0)}<\tilde{a}_{n+1}^{(0)}$ (the $n$-photon accuracy rises with number of photons), $p(\tilde{a}_1^{(0)})$ peaks between $\tilde{a}_n^{(0)}=0.5$ and $0.6$ as found by \cite{Tinsleyetal2016}, whereas $p(\tilde{a}_{>1}^{(0)})$ is broader due to our lack of knowledge about these accuracies. The $n_{\text{max}}$-contribution to the hyper-priors in equation (\ref{eqBetaScalePriors}) ensures that the distribution $p(\vec{\tilde{a}}^{(0)})$ is constant irrespective of the value of $n_{\text{max}}$. For the posterior we use the same transformations as described for the prior, hence $\tilde{a}_0^{(1)}=1/2$ and $\tilde{a}_n^{(1)}<\tilde{a}_{n+1}^{(1)}$ as required by common sense.\\
To determine the likelihood of $\vec{\Sigma}$ given $\vec{\tilde{a}}$ and $T$, we use the binomial distribution
\begin{equation}
	p\left(\vec{\Sigma}|T,\vec{\tilde{a}}\right)=\prod_{d=1}^{D}\binom{T}{\Sigma_d}\tilde{A}_d^{\Sigma_d}\left(1-\tilde{A}_d\right)^{T-\Sigma_d}
\end{equation}
where $\vec{\tilde{A}}=\vec{1}-\boldsymbol{\rho}(\vec{1}-\vec{\tilde{a}})$, see equation (\ref{eqAccuracyP}).\\

In order to determine $n_{\text{max}}$, we have to show that $\tilde{a}_{n_{\text{max}}+1}$ is irrelevant for the reconstruction, hence, that $p(\tilde{a}_{n_{\text{max}}+1}^{(0)})=p(\tilde{a}_{n_{\text{max}}+1}^{(1)})$ -- the prior equals the posterior, see appendix \ref{ssecnmaxproblem} for further details. The determination of $n_{\text{max}}$ starts with a reconstruction at a value of $n_{\text{max}}$ too low (typically $\bar{N}_{\text{max}}+1$). We add $\tilde{p}_{n_{\text{max}}+1}^{(0)}$ to $\vec{\tilde{p}}^{(0)}$ and multiplex the reconstruction. I.e., we perform the reconstruction $N_{\text{mult}}$ times in parallel by mapping $\vec{\tilde{p}}^{(0)}\mapsto\boldsymbol{\tilde{p}}^{(0)}$, a matrix of dimensions $(n_{\text{max}}+2)\times N_{\text{mult}}$. Accordingly, the vectors $\vec{\tilde{a}}^{(0)}$ and $\vec{\tilde{A}}^{(0)}$ are mapped to matrices $\boldsymbol{\tilde{a}}^{(0)}$ and $\boldsymbol{\tilde{A}}^{(0)}$ with dimensions $(n_{\text{max}}+2)\times N_{\text{mult}}$ and $D\times N_{\text{mult}}$ respectively. The $\boldsymbol{\rho}$-matrix is augmented to size $D\times (n_{\text{max}}+2)$ to incorporate the probability of sending $n_{\text{max}}+1$ photons.\\ 
The likelihood of the data $\vec{\Sigma}$ is evaluated separately for each of the multiplexes. For $\tilde{p}_{n_{\text{max}}+1}$ we set a $\text{beta}(2.5,0.5)$-prior which we transform to the domain $[p_{\text{min}},1]$, such that $\tilde{p}_{n_{\text{max}}+1}^{(0)}>\tilde{p}_{n_{\text{max}}}^{(0)}$ is likely. Initially, upon starting the $n_{\text{max}}$ determination we set $p_{\text{min}}=0$. We compare the $N_{\text{mult}}$ MCs of $\tilde{p}_{n_{\text{max}}+1}^{(0)}$ and $\tilde{p}_{n_{\text{max}}+1}^{(1)}$ based on which it is decided whether $n_{\text{max}}$ is set to a sufficiently high value (i.e., $p(\tilde{p}_{n_{\text{max}}+1}^{(0)})$ and $p(\tilde{p}_{n_{\text{max}}+1}^{(1)})$ are indiscernible). In case we find that $n_{\text{max}}$ is insufficient, we increase the value by $1$ and update the value of $p_{\text{min}}$ as described in appendix \ref{ssecnmaxproblem}. Then we perform the multiplexed reconstruction again until a sufficient value for $n_{\text{max}}$ has been determined.\\

The \textsc{RStan}-program returns MCs for all matrix elements of $\boldsymbol{\tilde{a}}^{(0)}$, $\boldsymbol{\tilde{a}}^{(1)}$ and $\boldsymbol{\tilde{A}}^{(1)}$. However, because $\boldsymbol{\tilde{a}}^{(1)}$ is the multiplexed version of $\vec{\tilde{a}}^{(1)}$ and all multiplexes have been compared to the same data, the MCs for all multiplexes can be appended together, i.e., we can ``squeeze'' the columns of $\boldsymbol{\tilde{a}}^{(0)}$, $\boldsymbol{\tilde{a}}^{(1)}$ and $\boldsymbol{\tilde{A}}^{(1)}$ back to $\vec{\tilde{a}}^{(0)}$, $\vec{\tilde{a}}^{(1)}$ and $\vec{\tilde{A}}^{(1)}$, such that the MC for each $\tilde{a}_n^{(0)}$, $\tilde{a}_n^{(1)}$ or $\tilde{A}_d$ contains $N_{\text{mult}}\cdot N_{\text{MC}}$ samples.\\
From the latter MCs we can determine the posterior statistics, such as means, medians, modes and high-density intervals (HDIs) using the \textsc{DBDA2E-utility.R}-functions provided by \cite{Kruschke2015}. \\
An example of the posterior statistics of $\vec{\tilde{a}}^{(1)}$ and $\vec{\tilde{A}}^{(1)}$ is discussed in the next section.

\section{A single simulation and reconstruction result}\label{secSingleSim}
\begin{figure}[tbhp]
	\centering
	\begin{minipage}[t]{0.50\textwidth}
		\includegraphics[width=0.99\textwidth]{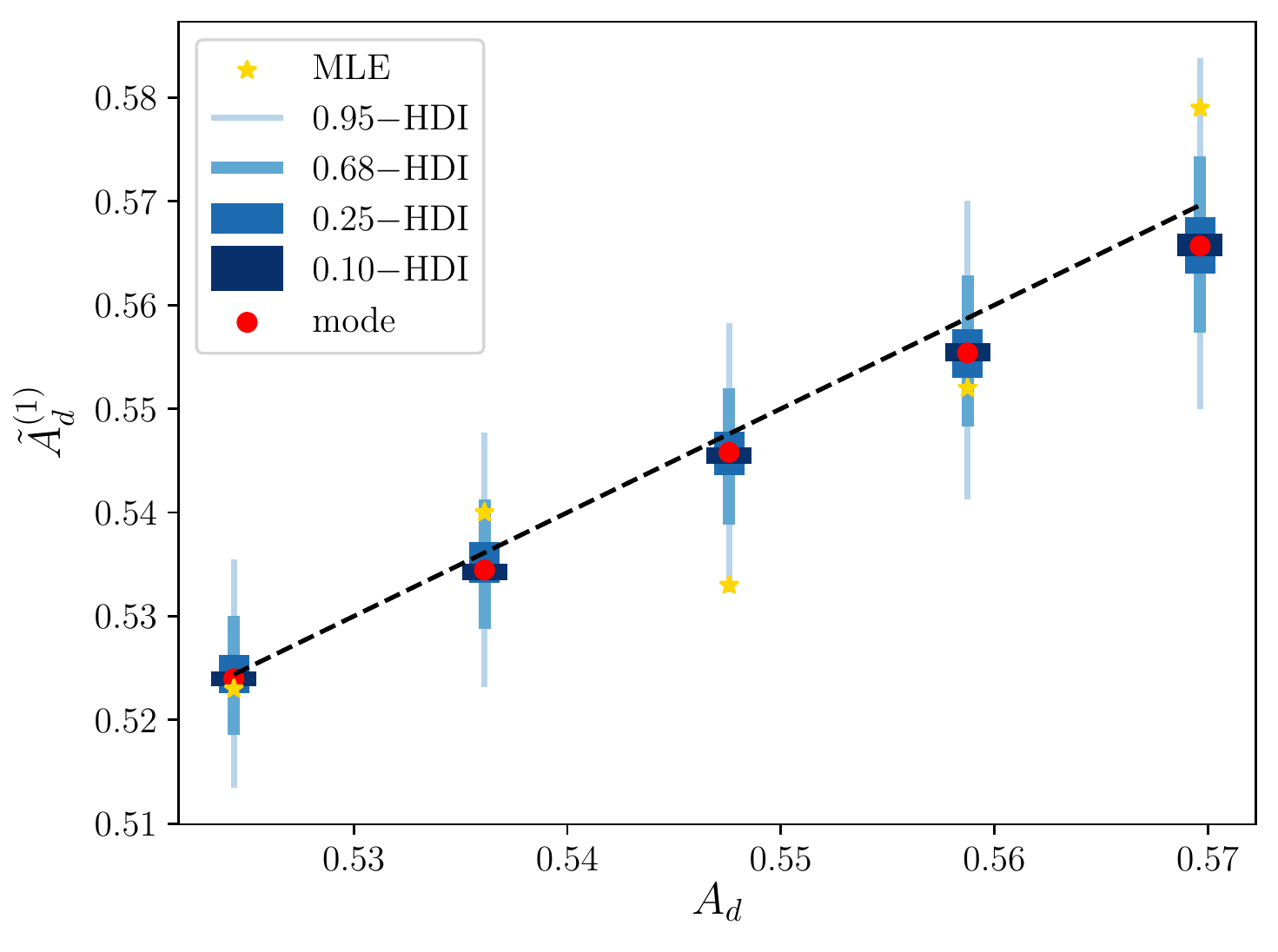}
		\caption{Reconstruction of the accuracy per data point (source intensity) for a single simulation. $\vec{A}$ are the model values obtained from substituting our model for $\vec{a}$ (equations (\ref{eqBinomModel}) and (\ref{eqptoa})) into equation (\ref{eqAccuracyP}). The black dashed line indicates $\tilde{A}_d^{(1)}=A_d$ to guide the eye. Although the most likely accuracy values ($A_d^{\text{MLE}}=\Sigma_d/T$, yellow asterisks) are different from the model values, the reconstruction $\vec{\tilde{A}}^{(1)}$ (indicated by specified HDI-intervals and the mode) is closer to the model values as a result of the detector tomography techniques we apply.} \label{fig_Arec_single}
	\end{minipage}\hfill
	\begin{minipage}[t]{0.49\textwidth}
		\includegraphics[width=0.99\textwidth]{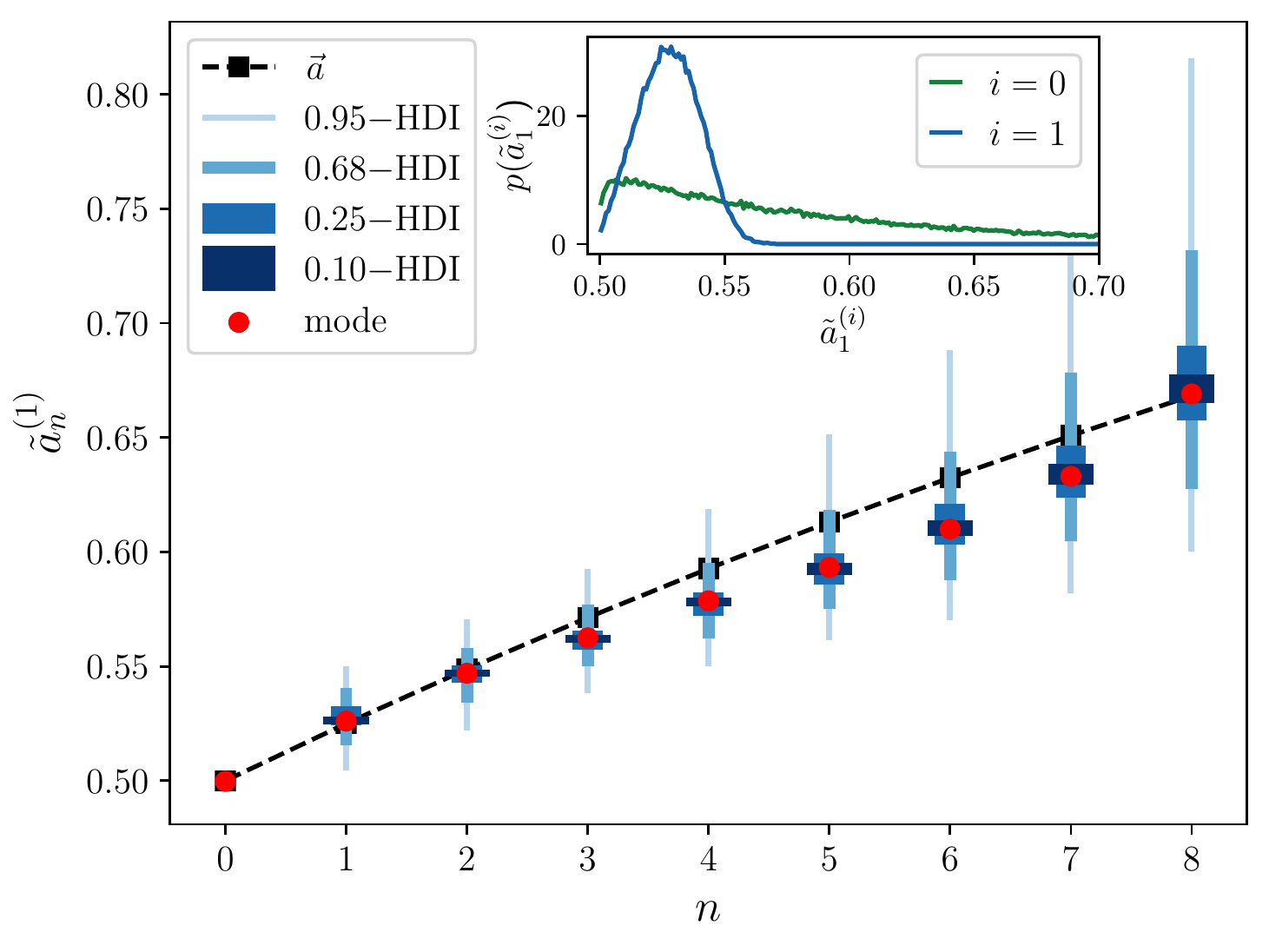}
		\caption{Reconstruction of the accuracy for $n$-photon states for a single simulation. The posterior distributions of $\tilde{a}_n^{(1)}$ are indicated using specified HDIs and the distribution mode. The black dashed line corresponds to our model values. The inset shows the prior and posterior distributions for $\tilde{a}_1$. From this figure one may estimate the Savage-Dickey ratio (see text), $r_{\text{SD}}$, at $\tilde{a}_1=0.5$ as indicator whether the data exclude $\tilde{a}_1^{(1)}=0.5$ (a single photon is undetected). For this simulation $r_{\text{SD}}(0.5)=-4.9$}\label{fig_arec_single}
	\end{minipage}
\end{figure}

Let us perform a single simulation as described in the previous section. We set $\bar{N}_{\text{min}}=1.0$, $\bar{N}_{\text{max}}=3.0$ and $D=5$ such that $\vec{\bar{N}}=[1.0,1.5,2.0,2.5,3.0]$, and $\sigma_{\bar{N},d}=0$ (no noise). For each data point we perform $T=1000$ trials, i.e., we run a total of $5000$ trials equally divided over $5$ data points. For the reconstruction we set $N_{\text{mult}}=7$, $N_{\text{chains}}=3$, $N_{\text{iter}}=15000$, $N_{\text{warmup}}=2500$ and $N_{\text{thin}}=3$. This yields $N_{\text{MC}}=12500$ per multiplex and therefore $87500$ MCMC samples per $\tilde{a}_n^{(1)}$ in total.\\
The reconstruction for the constant source accuracy $\vec{A}$, $\vec{\tilde{A}}^{(1)}$ is shown in figure \ref{fig_Arec_single}. Here, we plot the reconstructed values the experimental outcome $A_d^{\text{MLE}}=\Sigma_d/T$ with yellow asterisks. Indicated HDIs for each $\tilde{A}_d^{(1)}$ are spanned by blue lines, whereas the posterior modes of $\vec{\tilde{A}}^{(1)}$ are marked with a red dot giving an impression of the posterior distribution. As can be observed, the reconstructed values are close to the model values, which we obtain by direct substitution of the $n$-photon accuracies $\vec{a}$ (equations (\ref{eqBetaScalePriors}) and (\ref{eqptoa})) into equation (\ref{eqAccuracyP})). This is a direct result from the detector tomography technique we apply.\\
In figure \ref{fig_arec_single} the reconstructed values for $\vec{a}$, $\vec{\tilde{a}}^{(1)}$, can be observed, obtained from the same simulation. The posterior distributions corresponding to each $\tilde{a}_n^{(1)}$ are indicated in the same fashion as for $\tilde{A}_d^{(1)}$. We find good agreement between $\vec{a}$ and $\tilde{\vec{a}}^{(1)}$. Comparing figures \ref{fig_arec_single} to figure \ref{fig_Arec_single}, we see that $\vec{\tilde{a}}^{(1)}$ ``follows'' $\vec{\tilde{A}}^{(1)}$, as one should expect. Low (High) values for $A_d^{\text{MLE}}$ drags (lifts) $\vec{\tilde{a}}^{(1)}$ down (up).\\

From this reconstruction, we can determine whether single photons are detected ($\tilde{a}_1^{(1)}>0.5$). To this end, we estimate the Savage-Dickey ratio \cite{Dickey1971},
\begin{equation}
	r_{\text{SD}}\left(0.5\right)=10\log_{10}\left(\dfrac{p\left(\tilde{a}_1^{(1)}=0.5\right)}{p\left(\tilde{a}_1^{(0)}=0.5\right)}\right).
\end{equation}
This ratio allows us to exclude a value of $0.5$ for $\tilde{a}_1^{(1)}$, if the posterior drops significantly below the prior, i.e., in our case a \textit{de}creasing $r_{\text{SD}}\left(0.5\right)$ indicates \textit{in}creasing evidence that a single photon can indeed be detected. Although other decision criteria are available (see e.g. \cite{Kelter2020}), we choose to evaluate the Savage-Dickey ratio here, because of its ease in interpretation.\\
The prior and posterior distributions $p(\tilde{a}_1^{(0)})$ and $p(\tilde{a}_1^{(1)})$ for the simulation under consideration in this section have been depicted in the inset of figure \ref{fig_arec_single}. Fitting these distributions with a logspline-function from the \textsc{logspline}-library \cite{Rlogspline} available for \textsc{R} gives us an estimate for $r_{\text{SD}}(0.5)$. For this simulation we found $r_{\text{SD}}(0.5)=\SI{-4.9}{dB}$, which is on the verge of being considered substantial evidence that single photons can be detected \cite{Jeffreys1961}.\\
We can also estimate how well the higher-photon number accuracies are reconstructed. For this purpose we calculate the mean of squared errors of $\text{mode}(\vec{\tilde{a}}^{(1)})$ relative to $\vec{a}$,
	\begin{equation}
		\text{MSE}(a_n)=\dfrac{1}{n_{\text{max}}}\sum_n \left(\text{mode}(\tilde{a}_n^{(1)})-a_n\right)^2.
	\end{equation}
For the reconstruction result presented in figures \ref{fig_Arec_single} and \ref{fig_arec_single}, this calculation evaluates to $\SI{1.66e-4}{}$, implying the reconstructed mode is $1.29\%$ off on average for the $8$ reconstructed accuracies.

\section{Optimisation of experimental parameters for single-photon detection}\label{secOptimise}
Using the simulation and reconstruction algorithm, we can optimise the experimental parameters $\vec{\bar{N}}$, $D$ and $T$ in order to achieve a certain goal. Apart from that, we study the influence of the model parameters $p_1$ and $\sigma_{\bar{N},d}$. Being interested whether single photons are detected, we consider the Savage-Dickey ratio in this section and estimate the experimental success probability $p(r_{\text{SD}}(0.5)<r_{\text{SD},0})$, where $r_{\text{SD},0}$ is the threshold value below which we consider $\tilde{a}_1^{(1)}=0.5$ excluded. In appendix \ref{appOptimise} we consider some other optimisation goals.\\ 
$p(r_{\text{SD}}(0.5)<r_{\text{SD},0})$ is estimated by performing $100$ simulations for given experimental parameters and calculating the Savage-Dickey ratio for each. Since these simulations can be considered as binomial trials, in which $r_{\text{SD}}(0.5)<r_{\text{SD},0}$ implies success, we determine the uncertainty in the success probability by setting a $\text{beta}(1,1)$-prior (flat prior) for this parameter. Then, the posterior is a $\text{beta}(k+1,100+1)$-distribution, where $k$ is the number of simulations for which $r_{\text{SD}}(0.5)<r_{\text{SD},0}$.\\
\begin{figure}[t!]
	\centering
	\includegraphics[width=\textwidth]{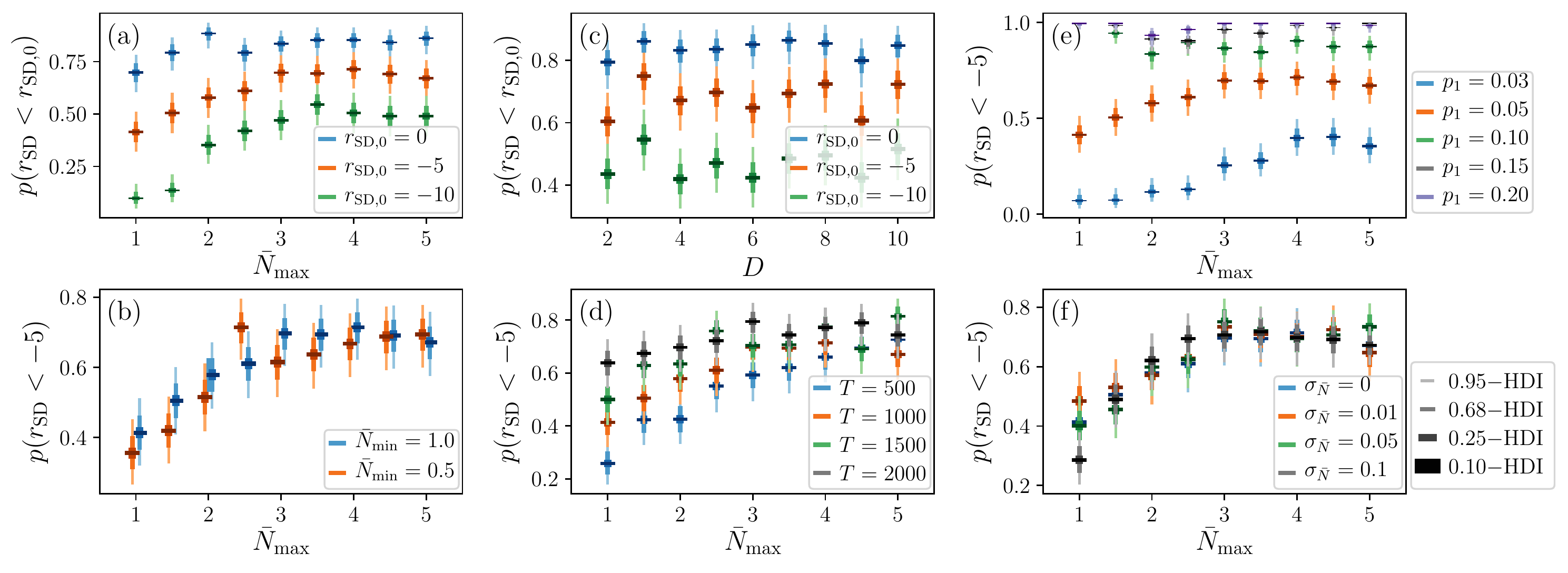}
	\caption{Experimental success probability based on the Savage-Dickey ratio, $p(r_{\text{SD}}(0.5)<r_{\text{SD},0})$, under variation of experimental and model parameters. (a) The experimental success probability as a function of $\bar{N}_{\text{max}}$ ($\bar{N}_{\text{min}}=1.0$, $T=1000$, $D=5$, $p_1=0.05$ and $\sigma_{\bar{N},d}=0$. MCMC parameters as stated in text.). The success probability rises linearly for low $\bar{N}_{\text{max}}$ and becomes constant for $\bar{N}_{\text{max}}>3$, irrespective of the cut-off ratio chosen for marking a successful experiment. (b) The value for $\bar{N}_{\text{min}}$ is of minor importance. For $\bar{N}_{\text{min}}=1.0$ and $0.5$ respectively (other parameters equal to the values in (a)), the maximum experimental success probability is hardly influenced by $\bar{N}_{\text{min}}$. The data are offset by $\pm0.05$ in their $\bar{N}_{\text{max}}$-value for clarity. (c) Varying $D$ while keeping $D\cdot T$ constant at $5000$ trials shows that also $D$ is of minor importance ($\bar{N}_{\text{max}}=3.0$ -- other parameters as stated under (a)). (d) Influence of the total number of trials by varying $T$ (other parameters as in (a)). It is observed that the slope in the experimental success probability for $\bar{N}_{\text{max}}<3.0$ decreases, while its constant value increases slightly with $T$. (e) Variation in the visual model value $p_1$ (all other parameters equal to those in (a)) reveals that the value of $\bar{N}_{\text{max}}$ for which the experimental success probability becomes constant decreases with $p_1$. (f) The reconstruction is resilient to noise in the light source. Varying $\sigma_{\bar{N}}$ (other parameters the same as in (a)), it is observed that the experimental success probability does not vary, except for $\bar{N}_{\text{max}}=1.0$.\\ 
	Based on these results, the optimal experimental parameters seem to be $\bar{N}_{\text{min}}=1.0$, $\bar{N}_{\text{max}}=4.0$, $D\geq 3$ and $D\cdot T\geq 5000$, while noise in the light source is not detrimental to the experiment.
	}\label{figSuccessOpt}
\end{figure}

Our optimisation results are presented in figure \ref{figSuccessOpt}. For $\bar{N}_{\text{max}}$ we consider values up to $5.0$, for which $\rho_1$, equation (\ref{eqPoisson}), has decreased to only $3.4\%$. On the other hand, since $\rho_1$ maximises for $\bar{N}=1$, we consider this value as a maximum for $\bar{N}_{\text{min}}$.\\ 
In general, we observe that the experimental success probability shows two regimes. For lower $\bar{N}_{\text{max}}$ the success probability is seen to rise linearly, whereas above a critical $\bar{N}_{\text{max,c}}$ it remains constant. This behaviour is apparent under variation of all other experimental and model parameters (except $D$), and $r_{\text{SD},0}$. As can be seen in figure \ref{figSuccessOpt}a, in which we plotted the success probability for $\bar{N}_{\text{min}}=1.0$, $p_1=0.05$, $D=5$, $T=1000$ and a noise-less source, $\bar{N}_{\text{max,c}}\approx 3.0$ does not depend on $r_{\text{SD},0}$. Taking $r_{\text{SD},0}=-5$, it is seen that $p(r_{\text{SD}}(0.5)<r_{\text{SD},0})$ rises to approximately $0.7$. It rises more or less to the same value if we set $\bar{N}_{\text{min}}=0.5$, as can be observed in figure \ref{figSuccessOpt}b. It reaches this value for slightly higher $\bar{N}_{\text{max}}$ and in general the success probability seems to be lower for $\bar{N}_{\text{min}}=0.5$ than for $\bar{N}_{\text{min}}=1.0$, except for $\bar{N}_{\text{max}}=2.5$. We attribute these results to the fact that $\rho_1$ maximises for $\bar{N}=1$, such that for any $\vec{\bar{N}}$ which contains $\bar{N}=1$ sending one photon is most likely. Therefore we infer that $\bar{N}_{\text{min}}=1$ will be optimal for our experiments.\\
In figure \ref{figSuccessOpt}c we plot our results for varying $D$, while keeping $D\cdot T=5000$ ($\bar{N}_{\text{min}}=1.0$, $\bar{N}_{\text{max}}=3.0$, other parameters the same). Thus, it is clearly observed that it does not matter if we perform many trials on a few source intensities, or rather a few trials for many source intensities. This can be understood since the amount of trials per photon number is more or less constant for every combination of $D$ and $T$ if $D\cdot T$ is kept constant. However, it should be noted that if one optimises for the smallest $\tilde{a}_1^{(1)}$ $0.95$-HDI, which is performed in appendix \ref{appOptimise}, $D\geq 3$ is a better choice. Hence, we would still advise to perform the experiment at at least $3$ source intensities.\\
The final experimental parameter we consider is the total number of trials $D\cdot T$. As can be seen in figure \ref{figSuccessOpt}d, $\bar{N}_{\text{max,c}}$ does not vary with $D\cdot T$ (setting $D=5$, other parameters as in figure \ref{figSuccessOpt}a). However, with increasing $D\cdot T$ the slope at $\bar{N}_{\text{max}}<\bar{N}_{\text{max,c}}$ decreases and the constant value which the success probability attains increases. The latter increase, however, is only small from which it follows that $5000$ trials ($T=1000$) is sufficient for the experiment.\\

Apart from the experimental parameters, we also study the influence of the model parameters $p_1$ and $\sigma_{\bar{N},d}$. In figure \ref{figSuccessOpt}e we plot our results for the experimental success probability while varying $p_1$ (other parameters the same as stated for figure \ref{figSuccessOpt}a). $p_1=0.03$ and $p_1=0.20$ correspond to the accuracies reported by \cite{Tinsleyetal2016} averaged over all trials and high-confidence trials respectively. It is clearly observed that the success probability drops while decreasing $p_1$. This is, of course, due to $a_1$ becoming closer to $0.5$, such that a difference from $0.5$ is harder to detect. Interestingly, we also observe that $\bar{N}_{\text{max,c}}$ increases with decreasing $p_1$. Based on this observation, we would advise setting $\bar{N}_{\text{max}}$ to $4.0$ during the actual experiment, because the actual value for $p_1$ not well-known.\\
Finally, we study the influence of noise on the light source. We set $\sigma_{\bar{N},d}$ to a constant $\sigma_{\bar{N}}$. Although such a model ($\sigma_{\bar{N},d}$ constant for all data points) would only be valid for light sources which are linear, we expect that this simple model reveals the most important aspects of a noisy light source. We observe in figure \ref{figSuccessOpt}f that the influence of noise on the experiment is minor. Only for $\bar{N}_{\text{min}}=\bar{N}_{\text{max}}=1.0$ noise seems to be of negative influence to the success probability. We suspect that this mainly results from the photon number distribution being fairly constant under influence of our noise model. This follows from the Taylor expansion of $\rho_n$,
\begin{equation}
\begin{aligned}
	\rho_n&=\exp\left(-\left(\bar{N}_{d,0}+\mathrm{d}\bar{N}\right)\right)\dfrac{\left(\bar{N}_{d,0}+\mathrm{d}\bar{N}\right)^n}{n!}\\
	&=\exp\left(-\bar{N}_{d,0}\right)\dfrac{\bar{N}_{d,0}^n}{n!}\left(1-\left(1-\dfrac{n}{\bar{N}_{d,0}}\right)\mathrm{d}\bar{N}+\left(\dfrac{1}{2}-\dfrac{n}{\bar{N}_{d,0}}+\dfrac{{n\choose 2}}{\bar{N}_{d,0}^2}\right)\mathrm{d}\bar{N}^2-\dots\right)\\
	&\approx\exp\left(-\bar{N}_{d,0}\right)\dfrac{\bar{N}_{d,0}^n}{n!}\left(1-\left(1-\dfrac{n}{\bar{N}_{d,0}}\right)\mathrm{d}\bar{N}\right)
\end{aligned}
\end{equation}
Upon direct comparison of the first and last line of this equation, we find that the approximation holds for $\mathrm{d}\bar{N}<0.15\bar{N}_{d,0}$. Since the approximation is linear in $\mathrm{d}\bar{N}$, the photon number distribution remains more or less constant for any distribution $\mathrm{d}\bar{N}$ symmetric around $0$. From this argument we expect that noise on $\bar{N}_{d,0}$ yields no problem for the experiment as long as its distribution is symmetric and the standard deviation of $\bar{N}_{d}$ is less than $0.05\bar{N}_{d,0}$.\\

In summary, from our results we find that the experimental success probability for detecting that $\tilde{a}_1^{(1)}>0.5$ is optimised for setting $\bar{N}_{\text{min}}=1.0$, $\bar{N}_{\text{max}}=4.0$, $D\geq 3$ and $D\cdot T\geq 5000$. The standard deviation of the noise on the light source should be less than $0.05\bar{N}_{d,0}$.

\section{Discussion}\label{secDiscussion}
Throughout this study we have made several assumptions, which influence on the experiments will be discussed below. Also, we will compare our approach of using a light source with a Poissonian photon number distribution to a source based on spontaneous parametric down conversion (SPDC), as used by \cite{Tinsleyetal2016}. Finally, we will describe how the data collected in the proposed experiment can also be used to determine the $n$-photon accuracies referenced to the retina. 

Since this is the first time that the method of detector tomography is considered for the human visual system, it was necessary to construct a model for $\vec{a}$. Under the assumption that the eye receptors function independently effectively, they receive at most one photon per light pulse and their quantum efficiency remains constant in time, we proposed a binomial model. Using a Maxwellian view, in which the light pulse is spread over a relatively large area of the retina (order $10^5$ receptors), it is likely that each receptor receives at most one photon. However, since visual signal processing is highly non-linear, we doubt that the assumption on effective independence of receptors can be made, even at this scale. Moreover, from \cite{Tinsleyetal2016} it is known that it is more likely to detect single photons if the time between single-photon events decreases. This implies the quantum efficiency of the receptors is not constant in reality. These points are even more valid for the case in which the light pulse is focussed on the retina, implying only order $10$ receptors partake in receiving the photons. In this case, one also cannot make the assumption any more that each photons lands on a different receptor. These effects can be studied experimentally by shifting the focal point of the light pulses between eye lens and retina and varying the time in between trials. Additionally, the irradiated part of the retina can be adapted by changing the focal distance of the focussing lens (the $f=\SI{50}{cm}$-lens discussed in section \ref{secEyeModel}). This would allow to study summation effects \cite{Holmesetal2017,Deyetal2021} in the visual system with few-photon number states.\\
We overcome the lack of knowledge on $\vec{a}$ by defining a prior that is able to take into account many models for $\vec{a}$, not one specific for our model. However, since we assume $\tilde{a}_{n+1}>\tilde{a}_n$, models for which $\tilde{a}_{n+1}\leq \tilde{a}_n$ are excluded. Also, although the amount of possible models that can be described using our prior is broad, not all possible models can be described. Therefore it remains a question whether a more suitable prior can be found.\\

Our other major assumption is that the test subject performs Bernoulli trials. In reality, however, $\vec{a}$ for a subject may vary in time (inter-session and intra-session) due to e.g. fluctuations in the quantum efficiency of the receptors (as already discussed), tiredness or other fluctuations in attention \cite{Schutt2016}. Therefore a beta-binomial likelihood, in which $\vec{a}$ is allowed to vary in between trials, might be more suitable. If the experiment shows that this is the case, \textsc{RStan} does allow to implement the beta-binomial likelihood easily. However, performing the reconstruction based on the binomial likelihood will return the mean $\vec{\tilde{a}}^{(1)}$, even if the beta-binomial likelihood proves to be more suitable. Another suggestion would be to perform the reconstruction with a hierarchical Bayesian model, in which the data is fitted per session and which also returns a hyper-$\vec{\tilde{a}}^{(1)}$ for the subject.\\
The same holds when considering different test subjects together. Because everyone's visual system is different one would expect that for every subject $\vec{a}$ differs as well. If the data of all subjects are pooled together, one would find the mean $\vec{\tilde{a}}^{(1)}$ for these subjects. Upon writing a hierarchical model, however, one would obtain information about each subject individually. It would even be possible to add two hierarchical layers, one inter-subject and another one inter-session.\\ 
If hierarchy is added to the reconstruction, the amount of trials rises accordingly. I.e., in case data of all participants is pooled, $5000$ trials is sufficient to obtain an estimate of their mean $\vec{\tilde{a}}^{(1)}$. When inter-test subject differences are taken into account, this number rises to $5000$ trials per participant, which is still a feasible number of trials. For studying intra-test subject effects, the number of trials would rise to $5000$ per session, which becomes unfeasible.\\ 

In terms of the question whether humans can detect single photons, it is interesting to compare our approach with the approach followed in \cite{Tinsleyetal2016} using a heralded single-photon source based on SPDC. In that work approximately $30000$ trials are performed of which about $2500$ are post-selected as single-photon trials. The study resulted in a confidence interval for the single-photon accuracy of $0.020$ (with a $p$-value of $0.0545$) around $a_1=0.516$. This value can be compared to our results, indicating a $\tilde{a}_1^{(1)}$ $0.95$-HDI length of around $0.048$ ($0.044$) for $5000$ ($10000$) trials, see appendix \ref{ssecOptNDT} for $\bar{N}_{\text{max}}=4.0$. Performing a single simulation and reconstruction using $30000$ trials ($\bar{N}_{\text{min}}=1.0$, $\bar{N}_{\text{max}}=4.0$, $p_1=0.05$, $D=5$, $T=6000$) we find that the $\tilde{a}_1^{(1)}$ $0.95$-HDI length drops to $0.036$, which is still significantly above $0.020$. This indicates that using an SPDC source yields more precise results for $\tilde{a}_1^{(1)}$. However, it would be interesting to apply detector tomography as described in this study to the data obtained in \cite{Tinsleyetal2016}, taking into account all trials, or by other authors.\\

It should be noted that the data from the proposed experiment are useful not only for determining $\vec{a}$ given $n$ photons are presented to the eye, but also for determining $\vec{a}_{\text{ret}}$, the accuracy that $n$ photons are detected given they are incident on the retina, similar to the work presented in \cite{Renemaetal2012}. To this end, one could model the cornea, eye lens and vitreous body as an absorber with transmission $\eta$ and map $\bar{N}\mapsto \eta\bar{N}$ in equation (\ref{eqAccuracyP}). Setting priors on $\eta$ and $\vec{a}_{\text{ret}}$ a reconstruction can be performed using the methods described in this work.

\section{Conclusions}\label{secConclusion}
We performed a feasibility study on applying detector tomography to the detection of few-photon number states by the human visual system. The main challenge in such an experiment is that the amount of trials is limited. Assuming a light source with Poissonian photon statistics, we simulated a $2$AFC experiment and reconstructed the photon number accuracies $\vec{a}$ for which we assumed a simple model. It was found that the reconstruction algorithm is able to reconstruct $\vec{a}$ well. Repeating simulations, we found the optimum experimental parameters to detect whether single photons are detectable (given our model): performing the experiment using at least $3$ source intensities bounded by a minimum average photon number of $1.0$ and a maximum average photon number of $4.0$, yields the highest chance of detecting $a_1$ to be different from $0.5$. Our results suggest that at least $5000$ trials are performed, equally distributed over the source intensities. The noise of the light source is expected not to be problematic for this experiment as long as the standard deviation of the average photon number of the source is less than $0.05\bar{N}_{d,0}$, $5\%$ of the nominal mean photon number per pulse. From this study, we conclude that detector tomography is a feasible technique to study human visual perception.\\
We note that the techniques explored in this study can be applied in a broader perspective. First, they are not limited by Poissonian statistics, but can in principle be applied using any light source as long as the photon statistics of the light presented to the eye are known. Furthermore, these techniques could possibly also be used to study other human senses, particularly the olfactory and gustatory system (Can humans smell/taste few-molecule states?). As such, quantum detector tomography opens the road to study human perception on the quantum level.

\subsection*{Acknowledgements}
This research was supported by the Dutch Research Council (NWO) with an NWO replication grant, \#$401.19.032$, to WL and YP.

\bibliographystyle{unsrt}
\bibliography{SPHVsims_arxiv.bbl} 

\begin{thebibliography}{10}

\bibitem{Lundeenetal2009}
J.S. Lundeen, A.~Feito, H.~Coldenstrodt-Ronge, K.L. Pregnell, Ch. Silberhorn,
  T.C. Ralph, J.~Eisert, M.B. Plenio, and I.A. Walmsley.
\newblock Tomography of quantum detectors.
\newblock {\em Nat. Phys.}, 5:27--30, 2009.

\bibitem{Feitoetal2009}
A.~Feito, J.S. Lundeen, H.~Coldenstrodt-Ronge, J.~Eisert, M.B. Plenio, and I.A.
  Walmsley.
\newblock Measuring measurement: theory and practice.
\newblock {\em New J. Phys.}, 11(9):093038, 2009.

\bibitem{Renemaetal2012}
J.J. Renema, G.~Frucci, Z.~Zhou, F.~Mattioli, A.~Gaggero, R.~Leoni, M.J.A.
  de~Dood, A.~Fiore, and M.P. van Exter.
\newblock Modified detector tomography technique applied to a superconducting
  multiphoton nanodetector.
\newblock {\em Opt. Express}, 20:2806--2813, 2012.

\bibitem{Loulakis2017}
M.~Loulakis, G.~Blatsios, C.S. Vrettou, and I.K. Kominis.
\newblock Quantum biometrics with retinal photon counting.
\newblock {\em Phys. Rev. Applied}, 8:044012, 2017.

\bibitem{Hechtetal1941}
S.~Hecht, S.~Schlaer, and M.H. Pirenne.
\newblock Energy at the threshold of vision.
\newblock {\em Science}, 93:585--587, 1941.

\bibitem{Sakitt1972}
B.~Sakitt.
\newblock Counting every quantum.
\newblock {\em J. Physiol.}, 223:131--150, 1972.

\bibitem{Teichetal1982}
M.C. Teich, P.R. Prucnal, G.~Vannucci, M.E. Breton, and W.J. McGill.
\newblock Multiplication noise in the human visual system at threshold: 1.
  quantum fluctuations and minimum detectable energy.
\newblock {\em J. Opt. Soc. Am.}, 72:419--431, 1982.

\bibitem{Tinsleyetal2016}
J.N. Tinsley, M.I. Molodtsov, R.~Prevedel, D.~Wartmann, J.~Espigul{\'e}-Pons,
  M.~Lauwers, and A.~Vaziri.
\newblock Direct detection of a single photon by humans.
\newblock {\em Nature Commun.}, 7:12172, 2016.

\bibitem{Barlow1977}
H.B. Barlow.
\newblock Retinal and central factors in human vision limited by noise.
\newblock In {\em Vertebrate photoreception}. Academic Press, London, 1977.

\bibitem{Hallett1987}
P.~E. Hallett.
\newblock Quantum efficiency of dark-adapted human vision.
\newblock {\em J. Opt. Soc. Am. A}, 4:2330--2335, 1987.

\bibitem{Donner1992}
K.~Donner.
\newblock Noise and the absolute thresholds of cone and rod vision.
\newblock {\em Vision Res.}, 32:853--866, 1992.

\bibitem{Wagenmakersetal2008}
E.-J. Wagenmakers, M.~D. Lee, T.~Lodewyckx, and G.~Iverson.
\newblock Bayesian versus frequentist inference.
\newblock In H.~Hoijtink, I.~Klugkist, and P.A. P.~A.~Boelen, editors, {\em
  Bayesian evaluation of informative hypotheses in psychology}, pages 181--207.
  Springer, 2008.

\bibitem{Cursioetal1990}
C.~Curcio, K.~Sloan, R.~Kalina, and A.~Hendrickson.
\newblock Human photoreceptor topography.
\newblock {\em J. Comp. Neurol.}, 292:497--523, 1990.

\bibitem{Holmesetal2017}
R.~Holmes, M.~Victora, R.F. Wang, and P.G. Kwiat.
\newblock Measuring temporal summation in visual detection with a single-photon
  source.
\newblock {\em Vision Res.}, 140:33--43, 2017.

\bibitem{Fechner1860}
G.T. Fechner.
\newblock {\em Elemente der psychophysik vol. 2}.
\newblock Breitkopf und H\"artel, Leipzig, Germany, 1860.

\bibitem{KingdomPrins2016}
F.A.A. Kingdom and N.~Prins.
\newblock {\em Psychophysics, a practical introduction}.
\newblock Academic Press, Cambridge, MA, $2^{\text{nd}}$ edition, 2016.

\bibitem{YeshurunCarrascoMaloney2008}
Y.~Yeshurun, M.~Carrasco, and L.T. Maloney.
\newblock Bias and sensitivity in two-interval forced choice procedures: Tests
  of the difference model.
\newblock {\em Vision Res.}, 48:1837--1851, 2008.

\bibitem{GarciaAlcala2011}
M.A. Garc{\'i}a-P{\'e}rez and R.~Alcal{\'a}-Quintana.
\newblock Interval bias in $2${AFC} detection tasks: sorting out the artifacts.
\newblock {\em Atten. Percept. Psychophys.}, 73:2332--2352, 2011.

\bibitem{Westheimer1966}
G.~Westheimer.
\newblock The maxwellian view.
\newblock {\em Vision Res.}, 6:669--682, 1966.

\bibitem{Rcore2017}
{R Core Team}.
\newblock {\em R: A Language and Environment for Statistical Computing}.
\newblock R Foundation for Statistical Computing, Vienna, Austria, 2017.

\bibitem{Rstan}
{Stan Development Team}.
\newblock {RStan}: the {R} interface to {Stan}, 2020.
\newblock R package version 2.21.2.

\bibitem{Kruschke2015}
J.~Kruschke.
\newblock {\em Doing bayesian data analysis}.
\newblock Academic Press, Boston, $2^{\text{nd}}$ edition, 2015.

\bibitem{Dickey1971}
James~M. Dickey.
\newblock {The Weighted Likelihood Ratio, Linear Hypotheses on Normal Location
  Parameters}.
\newblock {\em Ann. Math. Statist.}, 42:204 -- 223, 1971.

\bibitem{Kelter2020}
R.~Kelter.
\newblock How to choose between different bayesian posterior indices for
  hypothesis testing in practice.
\newblock \emph{e-print} arXiv:$2005.13181$, $2020$.

\bibitem{Rlogspline}
C.~Kooperberg.
\newblock {\em logspline: Routines for Logspline Density Estimation}, 2020.
\newblock R package version 2.1.16.

\bibitem{Jeffreys1961}
H.~Jeffreys.
\newblock {\em Theory of Probability}.
\newblock Oxford University Press, Oxford, UK, $3^{\text{rd}}$ edition, 1961.

\bibitem{Deyetal2021}
A.~Dey, A.J. Zele, B.~Feigl, and P.~Prakash~Adhikari.
\newblock Threshold vision under full-field stimulation: Revisiting the minimum
  number of quanta necessary to evoke a visual sensation.
\newblock {\em Vision Res.}, 180:1--10, 2021.

\bibitem{Schutt2016}
H.H. Sch\"utt, S.~Harmeling, J.H. Macke, and F.A. Wichmann.
\newblock Painfree and accurate bayesian estimation of psychometric functions
  for (potentially) overdispersed data.
\newblock {\em Vision Res.}, 122:105--123, 2016.

\end{thebibliography}

\appendix
\section{Reconstruction}\label{secReconstruction}
Using the data array with trial outcomes, we can try to reconstruct the model accuracies $\vec{a}=\left[a_0,a_1,\dots,a_{n_{\text{max}}}\right]$ set by the user. We perform the reconstruction under Bayesian inference. This implies that we have to supply an appropriate prior to the algorithm, discussed in section \ref{ssecPrior}. In section \ref{ssecnmaxproblem} we discuss a method to determine $n_{\text{max}}$ introduced in equation (\ref{eqAccuracyGen}).

\subsection{Prior considerations}\label{ssecPrior}
In order to construct an appropriate prior for our problem, we consider the following ``wish list'', where $\vec{a}$ refers to model (``real'') values and $\vec{\tilde{a}}$ refers to reconstructed values:
\begin{enumerate}
	\item No preference for explicit $\vec{\tilde{a}}$-model (e.g. the model presented in equations (\ref{eqBinomModel}) and (\ref{eqptoa})).
	\item $\tilde{a}_{n+1}>\tilde{a}_n$: It is more probable to accurately detect more photons.
	\item $\tilde{a}_0=1/2$: $0$ photons are undetectable by definition.
	\item $a_1\approx 0.516$ (\cite{Tinsleyetal2016}, all trials) or $a_1\approx 0.60$ (\cite{Tinsleyetal2016}, high-confidence trials): The prior for $\tilde{a}_1$ peaks in between to $0.5$ and $0.6$, and falls off for higher values.
	\item $\vec{\tilde{a}}_{>1}$: broader priors, and not ``too much'' conditional on $\tilde{a}_1$ (If $\tilde{a}_1=0.5$, then $\tilde{a}_{>1}$ should not necessarily be close to $0.5$).
	\item Priors on $\vec{a}$ should be independent on $n_{\text{max}}$.
\end{enumerate}
Instead of considering the prior on $\vec{a}$, $p(\vec{\tilde{a}}^{(0)})$, directly, we will first consider the prior on the detection probability $\vec{p}$, $p(\vec{\tilde{p}}^{(0)})$ (Again, the tilde refers to the value being reconstructed, and the $(0)$-superscript indicates a prior distribution. For posterior distributions we will use a $(1)$-superscript). The values of $\vec{\tilde{p}}^{(0)}$ can be easily transformed into values for $\vec{\tilde{a}}^{(0)}$ by equation (\ref{eqptoa}). Since these probabilities are bounded by $\left[0,1\right]$ a prior can be implemented by a beta-distribution, as will be discussed in the following.\\

\begin{figure}
	\centering
	\includegraphics[width=0.5\textwidth]{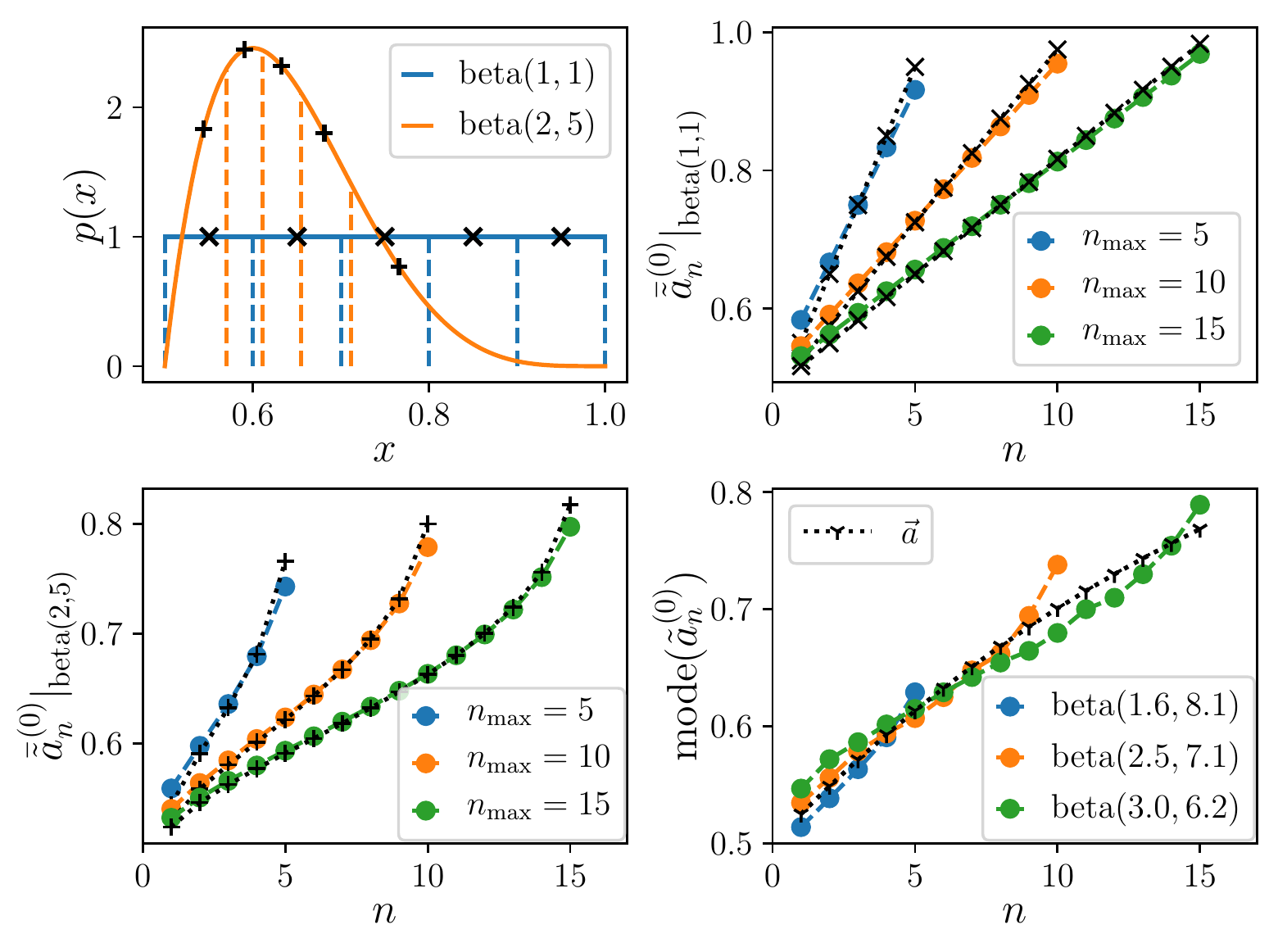}
	\caption{Understanding the beta$(\alpha,\beta)$-prior on $\vec{\tilde{a}}^{(0)}$. (top left) A simplified picture of the beta-prior on $\vec{\tilde{a}}^{(0)}$: If one subdivides the prior's distribution in $n_{\text{max}}$ domains of equal area, the mean of the $n^{\text{th}}$ domain corresponds to $\bar{\tilde{a}}_n^{(0)}$. Two examples are given for $n_{\text{max}}=5$: a beta$(1,1)$-prior (domain means indicated with `X') and a beta$(2,5)$-prior (domain means indicated with `+'). The same holds for the median (not depicted). (top right) The beta$(1,1)$-prior leads to a linear relation for $\bar{\tilde{a}}_n^{(0)}$. Depicted are the actual means (drawing and sorting values, colour) and the simplified means (as in top left panel, black) for several $n_{\text{max}}$. (bottom left) Same as (top right), but for the beta$(2,5)$-prior. (bottom right) The model described in equations (\ref{eqBinomModel}) and (\ref{eqptoa}) can be reasonably approximated by a beta-prior. The black curve indicates $\vec{a}$ (with $p_1=0.05$). The coloured curves show the modes of $\vec{\tilde{a}}^{(0)}$ from corresponding beta-priors approximating $a_n$ for different $n_{\text{max}}$. The shape parameters of these beta-priors have been determined by minimisation of the quantity $\sum(\text{mode}(\tilde{a}_n^{(0)})-a_n)^2$.}\label{figPriorUnderstanding}
\end{figure}

\begin{figure}[tbp]
	\centering
	\begin{subfigure}[b]{0.48\textwidth}
		\includegraphics[width=\textwidth]{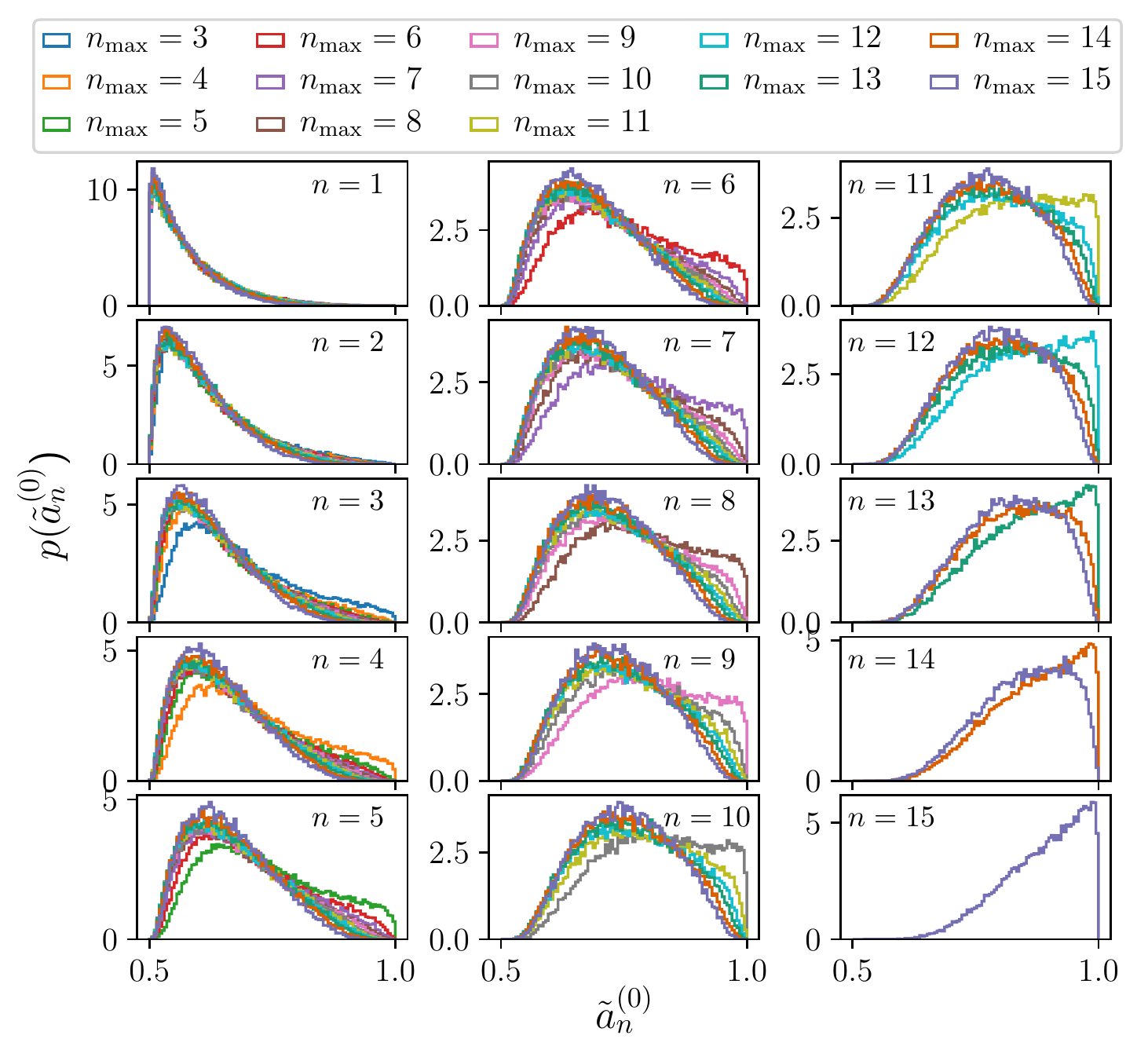}
		\caption{}\label{sfig_priorhists}
	\end{subfigure}
	\begin{subfigure}[b]{0.51\textwidth}
		\includegraphics[width=\textwidth]{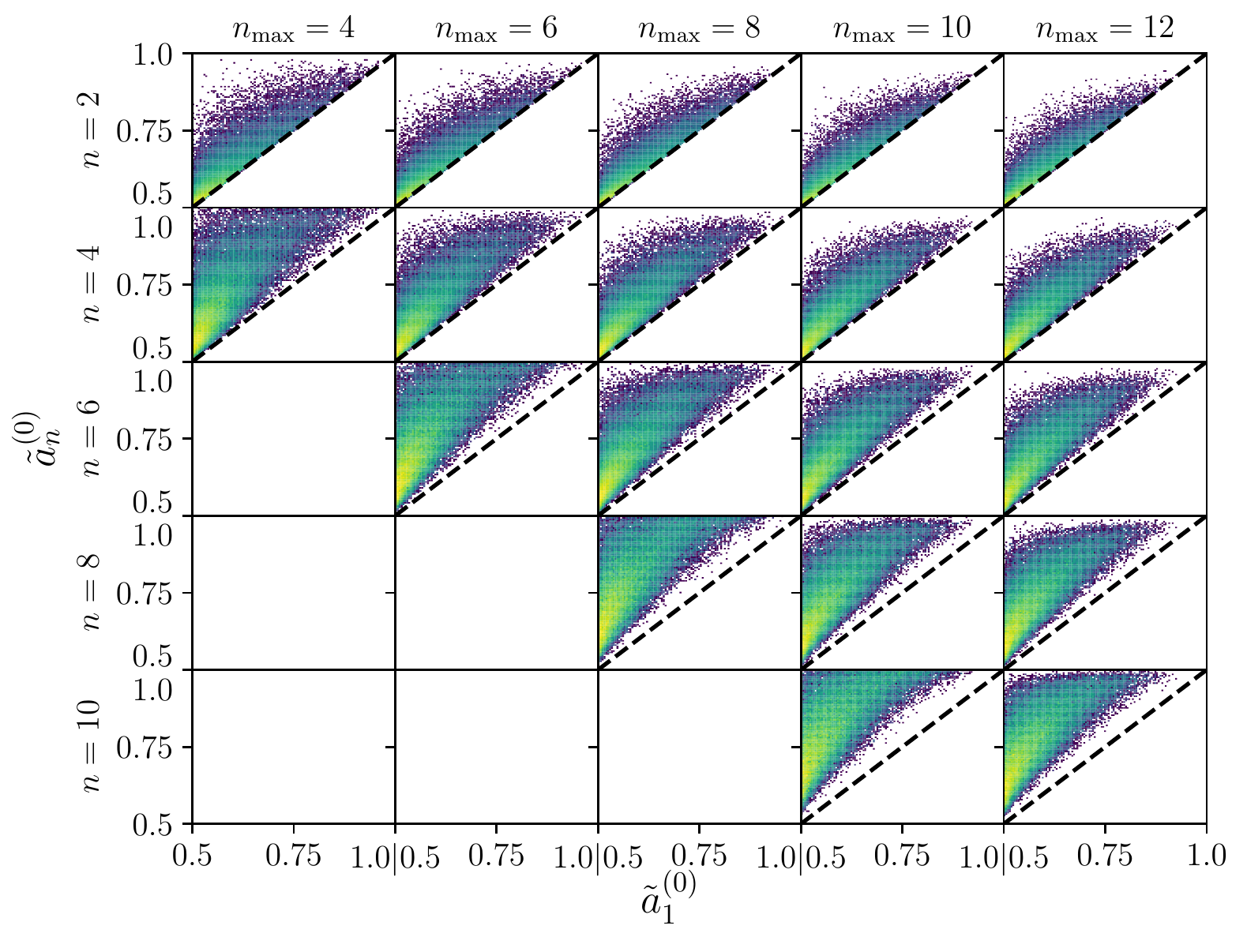}
		\caption{}\label{sfig_priorcorr}
	\end{subfigure}
	\caption{Prior on $\vec{\tilde{a}}^{(0)}$. (a) Prior distributions of $\tilde{a}_n^{(0)}$. Each of the subplots contains the indicated distributions for $n_{\text{max}}=3,4,\dots,15$, which are observed to comply with all items of our wish list, except item $5$ which compliance cannot be observed from this figure. (b) $2$-dimensional distributions (log-scale) featuring correlations between $\tilde{a}_1^{(0)}$ and $\tilde{a}_n^{(0)}$ for several $n$ and $n_{\text{max}}$. We see that $\tilde{a}_n^{(0)}$ is not conditional on $\tilde{a}_1^{(0)}$, as required by item $5$ on our wish list.}\label{figPrior}
\end{figure}

Suppose one chooses a beta$(1,1)$-prior for $\vec{\tilde{p}}^{(0)}$ and one would like to reconstruct data using $\tilde{a}_0,\tilde{a}_1,\dots,\tilde{a}_{n_{\text{max}}}$. Then, wish list conditions $2$ and $3$ can be enforced by drawing $n_{\text{max}}$ values from a $\text{beta}(1,1)$-distribution, sorting them and appending them to $0$. The resulting vector $\vec{\tilde{p}}^{(0)}=[0,\tilde{p}_1^{(0)},\dots,\tilde{p}_{n_{\text{max}}}^{(0)}]$ (where $\tilde{p}_n^{(0)}<\tilde{p}_{n+1}^{(0)}$) is then transformed to the prior $n$-photon accuracy vector by equation (\ref{eqptoa}).\\
It is important to realise that here the beta$(1,1)$-prior determines the prior function for $\vec{\tilde{p}}^{(0)}$ and thus $\vec{\tilde{a}}^{(0)}$. As can be observed in figure \ref{figPriorUnderstanding}, the uniform beta$(1,1)$-prior leads to a linear function for $\bar{\vec{\tilde{a}}}^{(0)}$. Different prior functions can be implemented by choosing different shape parameters for the beta-prior, as can be seen in the same figure. This covers condition $1$ from our wish list.\\
In order to provide some further understanding for the beta-prior, suppose one sets a beta$(\alpha,\beta)$-prior for $\vec{\tilde{p}}^{(0)}$. Insight in the resulting function follows from dividing the distribution of the beta-prior into $n_{\text{max}}$ equal sub-domains, as indicated in figure \ref{figPriorUnderstanding}, top left panel. Now the mean of each sub-domain corresponds to the mean of the $n_{\text{max}}$ $\tilde{p}_n^{(0)}$s, as can be seen in figure \ref{figPriorUnderstanding}, top right and bottom left panel for $\tilde{a}_n^{(0)}$. The same holds for the median (not depicted). Although the correspondence is not perfect, one can understand the prior function for $\vec{\tilde{p}}^{(0)}$ from this simple picture.\\ 
The other conditions ($4$--$6$) on our wish list can be covered by setting hyper-priors on the shape parameters of the beta-prior, $\alpha$ and $\beta$. By trial and error, we found that
\begin{align}\label{eqBetaScalePriors_app}
	\alpha&\sim\mathrm{halfnorm}\left(1,1/2+n_{\text{max}}/2-n_{\text{max}}^2/35\right),\\
	\beta&\sim\mathrm{halfnorm}\left(1,12-n_{\text{max}}/1.8-n_{\text{max}}^2/200\right)
\end{align}
covers these conditions, as can be seen in figure \ref{figPrior}. Here the halfnorm$(\mu,\sigma)$-distribution represents the half-normal distribution with mean $\mu$ and standard deviation $\sigma$.

\subsection{$n_{\text{max}}$-problem}\label{ssecnmaxproblem}
As discussed, we need to determine $n_{\text{max}}$, the number of photons at which the sum in equation (\ref{eqAccuracyP}) terminates. Termination may be either due to $a_{n}=1$ or $\rho_{n}=0$ for some $n$. Since our experiments will be run at relatively low $\bar{N}_{d}$, the sum will likely be terminated by the second condition. This we refer to as the $n_{\text{max}}$-problem: what is the maximum relevant number of photons to be taken into account in the reconstruction?\\
To illustrate, if $\bar{N}_d=1$, the probability to send $4$ photons equals $0.015$, hence for $1000$ trials we may expect $15$ trials with $4$ photons. For $5$ photons this number drops to $3$ and for $6$ it is $0.5$.\\

$n_{\text{max}}$ can be identified considering the first \emph{irrelevant} $n$, $n_{\text{max}+1}$. This number of photons will no longer influence the reconstruction, and as a result the prior for $\tilde{p}_{n_{\text{max}+1}}$ equals its posterior.\\
To this end we add $\tilde{p}_{n_{\text{max}}+1}^{(0)}$ to $\vec{\tilde{p}}^{(0)}$ and give it a prior distribution. In principle, this may be any prior on the domain $\left[p_{\text{min}},1\right]$, where $p_{\text{min}}$ needs to be set in order to (approximately) comply with the $\tilde{a}_{n+1}>\tilde{a}_n$-condition described in section \ref{ssecPrior}. We will discuss the parameter $p_{\text{min}}$ further towards the end of this section. For the prior on $\tilde{p}_{n_{\text{max}}+1}^{(0)}$ we choose the beta$(2.5,0.5)$-distribution, transformed to the appropriate domain, i.e., $\tilde{p}_{n_{\text{max}}+1}^{(0)}=p_{\text{min}}+(1-p_{\text{min}})\tilde{p}_{n_{\text{max}}+1,0}^{(0)}$, where $\tilde{p}_{n_{\text{max}}+1,0}^{(0)}\sim\text{beta}(2.5,0.5)$. It should be noted that the $\boldsymbol{\rho}$-matrix containing the theoretical probabilities for presenting $n$ photons to the subject now has dimensions $D\times\left(n_{\text{max}}+2\right)$.\\

\begin{figure}[tbhp]
	\centering
	\begin{minipage}[t]{0.50\textwidth}
		\includegraphics[width=0.99\textwidth]{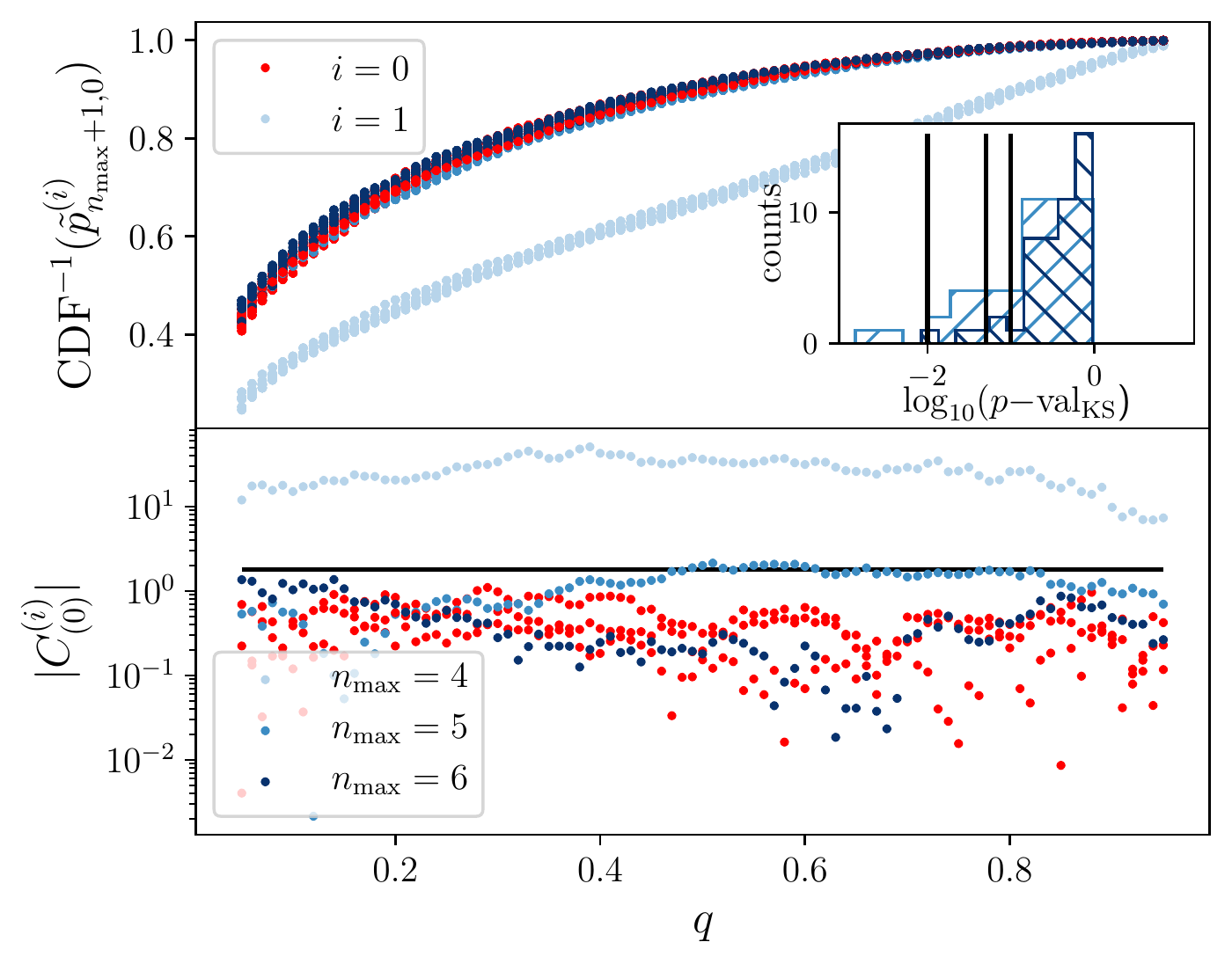}
		\caption{Determining $n_{\text{max}}$ for the simulation result presented in section \ref{secSingleSim} ($N_{\text{iter}}=5000$, $N_{\text{warmup}}=2500$, $N_{\text{chains}}=N_{\text{thin}}=3$, $N_{\text{mult}}=7$, $\tilde{p}_{n_{\text{max}}+1,0}^{(0)}\sim\text{beta}(2.5,0.5)$. NB: here we show only the data for low $N_{\text{iter}}=5000$, not those for high $N_{\text{iter}}=15000$ (see below) for which $n_{\text{max}}=8$ as seen in figure \ref{fig_arec_single}.). The legends hold for both panels. (top) prior (red) and posterior (blue) inverse CDF for $\tilde{p}_{n_{\text{max}}+1,0}^{(0)}$ for increasing $n_{\text{max}}$, showing $N_{\text{mult}}=7$ estimations for the curve. Upon increasing $n_{\text{max}}$, the prior and posterior converge. The inset shows a histogram of the $N_{\text{mult}}^2$ $p$-values returned by \textsc{R}'s KS test. The three lines indicate $p$-values of $0.01$, $0.05$ and $0.10$, which are often used in significance tests. As can be seen, $p$-values for $n_{\text{max}}=5$ and $6$ vary over a large range, such that they are not suitable to determine the equivalence of prior and posterior. (bottom) Instead, we use the $N_{\text{mult}}$ estimations to derive a parameter $C$ (see text), which equals $0$ if prior and posterior are equivalent. As can be observed, increasing $n_{\text{max}}$ lowers $|C_{(0)}^{(1)}|$. If $|C_{(0)}^{(1)}|<1.8$ (horizontal black line), we conclude we have reached the proper $n_{\text{max}}$ for the current reconstruction (see figure \ref{figCompval}).} \label{fig_p0andC}
	\end{minipage}\hfill
	\begin{minipage}[t]{0.49\textwidth}
		\includegraphics[width=0.99\textwidth]{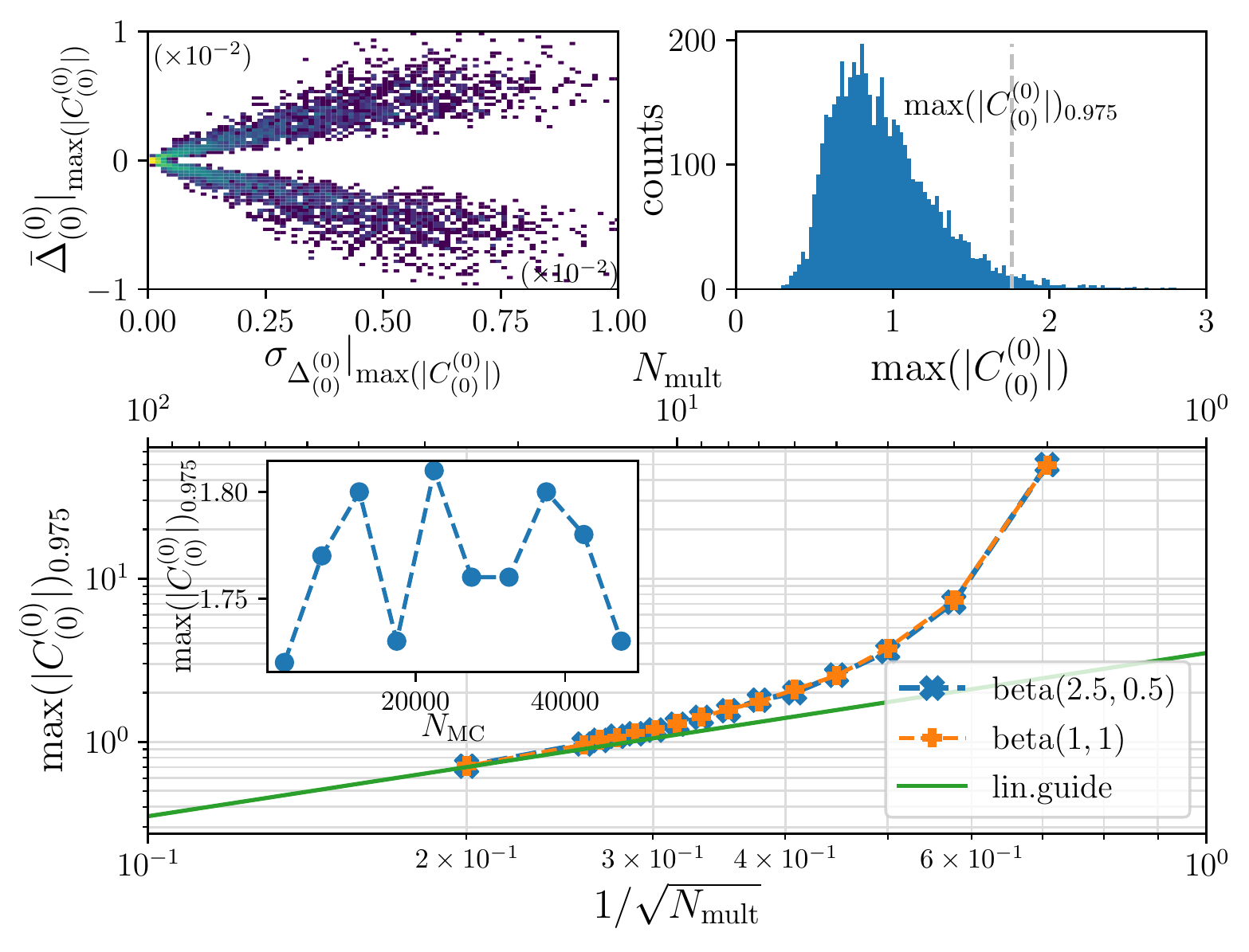}
		\caption{Determination of the cut-off value for $|C_{(0)}^{(1)}|$ . (top left) For each of $5000$ Monte Carlo simulations (same parameters as in figure \ref{fig_p0andC}) we determine $\bar{\Delta}_{(0)}^{(0)}$ and $\sigma_{\Delta_{(0)}^{(0)}}$ at the quantile maximising $|C_{(0)}^{(0)}|$. This yields a symmetric distribution around $\bar{\Delta}_{(0)}^{(0)}=0$. (top right) Histogram for the maximum values of $|C_{(0)}^{(0)}|$ from the Monte Carlo simulations. We determine the cut-off value of $|C_{(0)}^{(1)}|$ at the $0.975$-quantile of the corresponding distribution (indicated by the dashed line), which corresponds to a significance test at $p\text{-val}=0.05$-level. (bottom) The cut-off value for $|C_{(0)}^{(1)}|$ varies with $N_{\text{mult}}$ ($N_{\text{MC}}=12500$, as before). Upon increasing $N_{\text{mult}}$, the cut-off value tends to $\sqrt{N_{\text{mult}}}$-behaviour. Performing the same analysis for a beta$(1,1)$-prior, it seems the cut-off value is independent of the explicit (beta-)prior on $\tilde{p}_{n_{\text{max}}+1,0}^{(0)}$. The cut-off value is also independent of $N_{\text{MC}}$, as can be seen in the inset ($N_{\text{mult}}=7$, as before). For the parameters used in figure \ref{fig_p0andC}, $1.8$ is a suitable cut-off value.}\label{figCompval}
	\end{minipage}
\end{figure}

In order to test the equivalence of the prior of $\tilde{p}_{n_{\text{max}}+1}$ and its posterior, generally the Kolmogorov-Smirnoff (KS) test is applied. However, we apply a different method, since we found that the KS test puts too loose a bound on ``equivalence'' for our purpose. Therefore this test is likely to return an $n_{\text{max}}$ too low. This is illustrated in the inset of the top panel in figure \ref{fig_p0andC}. Upon inspection of equation (\ref{eqAccuracyP}) one finds that setting $n_{\text{max}}$ too low will lead to lower values for $\vec{\tilde{a}}$, because from a mathematical point of view, truncating the sum at $n_{\text{max}}$ is the same as assuming $\vec{a}_{>n_{\text{max}}}=1$. If relevant $\vec{\tilde{a}}_{>n_{\text{max}}}$ are set to $1$, which is too high, the reconstruction algorithm must choose lower values for $\vec{\tilde{a}}_{\leq n_{\text{max}}}^{(1)}$ to make up for the difference.\\
Instead of performing a KS test, we directly compare the prior and posterior inverse Cumulative Distribution Function (CDF) of $p_{n_{\text{max}}+1}$ by evaluating their difference, defining
\begin{equation}\label{eqDelta}
	\Delta_{(0)}^{(1)}\left(q\right)=\mathrm{CDF}^{-1}(\tilde{p}_{n_{\text{max}+1}}^{(0)})\left(q\right)-\mathrm{CDF}^{-1}(\tilde{p}_{n_{\text{max}+1}}^{(1)})\left(q\right).
\end{equation}
Here, the super- and subscripts $(0)$ and $(1)$ refer to the prior and posterior distribution as before. $q$ denotes the evaluated quantile running from $0$ to $1$. The inverse CDF for quantile $q$ can be approximated from the MC of some parameter by sorting the chain in ascending order, and then taking its $q\cdot\text{length(MC)}$-th element, which is rounded to yield an integer.\\
To take into account the intrinsic noise in MCs, we multiplex the reconstruction. I.e., instead of creating a single MC per $\tilde{a}_n^{(0)}$ and $\tilde{a}_n^{(1)}$, we create $N_{\text{mult}}$ MCs in parallel. This can be achieved by evaluating the different chains set by $N_{\text{chains}}$ separately, or, if one uses a computer with a low number of cores, by building in the multiplexing factor by hand. This is done by promoting the $(n_{\text{max}}+2)$-dimensional $\vec{\tilde{p}}^{(0)}$ to a matrix $\boldsymbol{\tilde{p}}^{(0)}$ with dimensions $(n_{\text{max}}+2)\times N_{\text{mult}}$. Accordingly, the vectors $\vec{\tilde{a}}^{(0)}$ and $\vec{\tilde{A}}^{(0)}$ are updated to matrices $\boldsymbol{\tilde{a}}^{(0)}$ and $\boldsymbol{\tilde{A}}^{(0)}$ with dimensions $(n_{\text{max}}+2)\times N_{\text{mult}}$ and $D\times N_{\text{mult}}$ respectively. Finally, the likelihood of the data is evaluated separately for each of the multiplexes.

Multiplexing the reconstruction $N_{\text{mult}}$ times for prior and posterior gives us $N_{\text{mult}}^2$ approximations for $\Delta_{(0)}^{(1)}(q)$. From these approximations we can estimate the quantity
\begin{equation}
	C_{(0)}^{(1)}(q)=\dfrac{\bar{\Delta}_{(0)}^{(1)}(q)}{\sigma_{\Delta_{(0)}^{(1)}(q)}}
\end{equation}
which measures how many standard deviations ($\sigma_{\Delta_{(0)}^{(1)}(q)}$) $\bar{\Delta}_{(0)}^{(1)}$ differs from $0$. Of course, $C_{(0)}^{(1)}(q)=0$ implies that the prior and posterior distributions are the same.\\
Figure \ref{fig_p0andC} plots $|C_{(0)}^{(1)}(q)|$ for the reconstruction presented in section \ref{secSingleSim}, while increasing $n_{\text{max}}$ from $4$ to $6$. In the top panel one observes the $N_{\text{mult}}=7$ prior and posterior inverse CDFs for $p_{n_{\text{max}}+1,0}$ (which has a beta$(2.5,0.5)$-prior). Additionally, we plot $|C_{(0)}^{(0)}(q)|$, which is defined similarly as $C_{(0)}^{(1)}(q)$, but compares a set of $N_{\text{mult}}$ different prior MCs to another set of $N_{\text{mult}}$ different prior MCs. As can be seen, $|C_{(0)}^{(1)}(q)|$ decreases with $n_{\text{max}}$, whereas $|C_{(0)}^{(0)}(q)|$ remains constant.\\

What remains is to set a cut-off on $|C_{(0)}^{(1)}(q)|$, on basis of which it is decided that the prior and posterior are different or equivalent. Similarly to the KS test, we consider $\max(|C_{(0)}^{(1)}(q)|)$ for determining such a cut-off, which may be dependent on the prior distribution considered for $\tilde{p}_{n_{\text{max}}+1,0}^{(0)}$, $N_{\text{mult}}$ and $N_{\text{MC}}$.\\
Performing a Monte Carlo simulation, we found that a critical value of $1.8$ is appropriate for the parameters used in figure \ref{fig_p0andC}, as can be seen in figure \ref{figCompval}. I.e., if
\begin{equation}\label{eqCcond}
	\max\left(C_{(0)}^{(1)}(q)\right)<1.8
\end{equation}
the prior and posterior are marked as equivalent. This value corresponds to the $0.975$-quantile of the $\max(|C_{(0)}^{(0)}(q)|)$-distribution, which is equivalent a $p$-value of $0.05$. Remarkably, this cut-off value does not depend on $N_{\text{MC}}$, nor does it seem to depend on the (beta-)prior on $p_{n_{\text{max}}+1,0}$. The cut-off value does depend on $N_{\text{mult}}$, as observed in figure \ref{figCompval}. For larger $N_{\text{mult}}$, the cut-off value tends to show $\sqrt{N_{\text{mult}}}$-behaviour.\\

In practice, we detect $n_{\text{max}}$ for a a single simulation in two stages and set $p_{\text{min}}$, introduced as the minimum value for $\tilde{p}_{n_{\text{max}}+1}$ at the beginning of this section, simultaneously. Fixing $N_{\text{chains}}$ (in this study typically $3$), $N_{\text{warmup}}$ (in this study typically $2500$) and $N_{\text{thin}}$ (in this study typically $3$), we set $N_{\text{mult}}$ (in this study typically $7$) and start a reconstruction at too low $n_{\text{max}}$ (typically $\bar{N}_{\text{max}}+1$), at low $N_{\text{iter}}$ (in this study typically $5000$) and with $p_{\text{min}}=0$. If condition (\ref{eqCcond}) is not met after performing a reconstruction with these parameters, we set $p_{\text{min}}=\text{mode}(\tilde{a}_{n_{\text{max}}}^{(1)})$ and we increase $n_{\text{max}}$ by $1$. Then we perform another reconstruction with these new parameters and continue until condition (\ref{eqCcond}) is met. This ensures that too low values for $n_{\text{max}}$ are quickly discarded and that $\tilde{p}_{n_{\text{max}}+1}^{(0)}$ is likely in excess of $\tilde{p}_{n_{\text{max}}}^{(0)}$.\\ 
Once condition (\ref{eqCcond}) is met, $N_{\text{iter}}$ is increased to a high value (in this study typically $15000$) and we continue the analysis. While switching from low to high $N_{\text{iter}}$ we do not adapt $n_{\text{max}}$ and $p_{\text{min}}$. If condition (\ref{eqCcond}) is met at $N_{\text{iter}}^{\text{high}}$, we conclude that $n_{\text{max}}$ has been detected.\\

\begin{figure}
	\centering
	\includegraphics[width=0.5\textwidth]{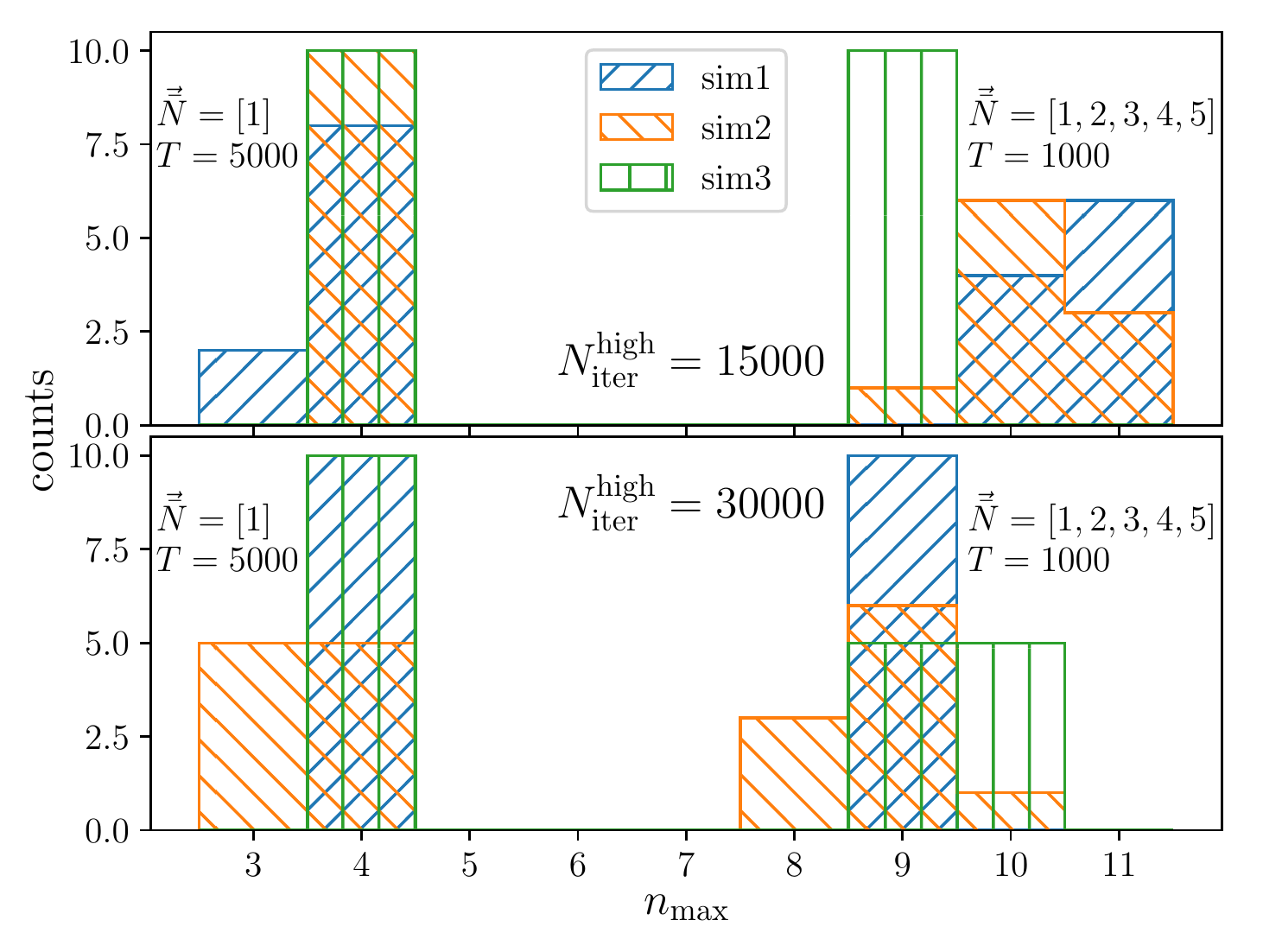}
	\caption{Outcomes of the $n_{\text{max}}$ determination for $N_{\text{iter}}^{\text{high}}=15000$ (top) and $N_{\text{iter}}^{\text{high}}=30000$ (bottom). On the left side of each panel, the results for $3$ simulations with $\vec{\bar{N}}=[1]$ and $T=5000$ is plotted, whereas the right side shows the same for $\vec{\bar{N}}=[1,2,3,4,5]$ and $T=1000$. Other relevant parameters have the typical values stated in text. Different reconstructions with the same data may lead to different values of $n_{\text{max}}$ due to noise in the MCs.}\label{fig_nmax_for_reps}
\end{figure}
This method does not ensure that the same $n_{\text{max}}$ is found for different reconstructions of the same data. Typically, $1$ to $3$ adjoining values are found, depending (mainly) the simulation result. This can be observed in figure \ref{fig_nmax_for_reps} and results from to the noise in the MCs. This persisted for of $N_{\text{iter}}^{\text{high}}$ up to $30000$ at least. Possibly, if $N_{\text{iter}}$ is increased further, the result of $n_{\text{max}}$ for a set of data differs less. Another possibility is to accept the different $n_{\text{max}}$ outcomes, and to combine different reconstruction results with the same data in order to get better approximations for $\vec{a}$. However, for the data presented in the following, we did not perform such a combination, due to the required computation time.

\section{Optimisation of experimental parameters for other figures of merit}\label{appOptimise}
With the simulation and reconstruction algorithm at hand, we can determine the optimum parameters for an experiment. In the main text we optimised the probability of excluding imperceptibility of single photons using the Savage-Dickey ratio. However, one can also optimise for other figures of merit. The parameters that can be influenced are $\vec{\bar{N}}$, $D$ and $T$. Apart from that, we will study the influence of our detection model and noise of the light source on the optimal experimental parameters. Firstly, however, we will discuss some figures of merit that we considered for optimising.

\subsection{Figures of merit}\label{ssecFigMerit}
We consider the following figures of merit to optimise the experimental parameters
\begin{itemize}
	\item The probability that the experiment yields an $r_{\text{SD}}(0.5)< r_{\text{SD},0}$, $p(r_{\text{SD}}(0.5)< r_{\text{SD},0})$. For example, $r_{\text{SD},0}=-5$ (substantial evidence that single photons are detected \cite{Jeffreys1961}) or $r_{\text{SD},0}=-10$ (strong evidence). The larger this probability, the more likely the experiment is to draw significant conclusions on single photon detection by humans.
	\item The mean and standard deviation of the $\tilde{a}_1^{(1)}$ $x$-HDI length, e.g. the $0.95$-HDI. The smaller an $x$-HDI length, the more precise we determine $\tilde{a}_1^{(1)}$.
	\item The probability that $a_1$ is in the $x$-HDI of $\tilde{a}_1^{(1)}$, $p(a_1\in \tilde{a}_1^{(1)} x\text{-HDI})$. Increasing this probability implies the reconstruction is more likely to return the model value for $a_1$.
	\item The mean of $\text{MSE}(a_n)$. This yields an indication how well higher-photon number accuracies are reconstructed.
\end{itemize}
We can estimate these parameters by repeating the simulation for specific  experimental settings ($\vec{\bar{N}}$, $D$ and $T$) and model parameters ($p_1$ and $\sigma_{\bar{N},d}$). To this end, we perform perform $100$ simulations as described in section \ref{secSingleSim} with the same experimental settings. To illustrate, we performed $100$ simulations with the same settings as in section \ref{secSingleSim}, i.e., $p_1=0.05$, $\vec{\bar{N}}=[1.0,1.5,2.0,2.5,3.0]$ and $T=1000$. For the reconstruction also use the same parameters $N_{\text{mult}}=7$, $N_{\text{chains}}=3$, $N_{\text{iter}}^{\text{low/high}}=5000/15000$, $N_{\text{warmup}}=2500$ and $N_{\text{thin}}=3$.\\ 

\begin{figure}[tbp]
	\centering
	\begin{subfigure}[b]{0.495\textwidth}
		\includegraphics[width=\textwidth]{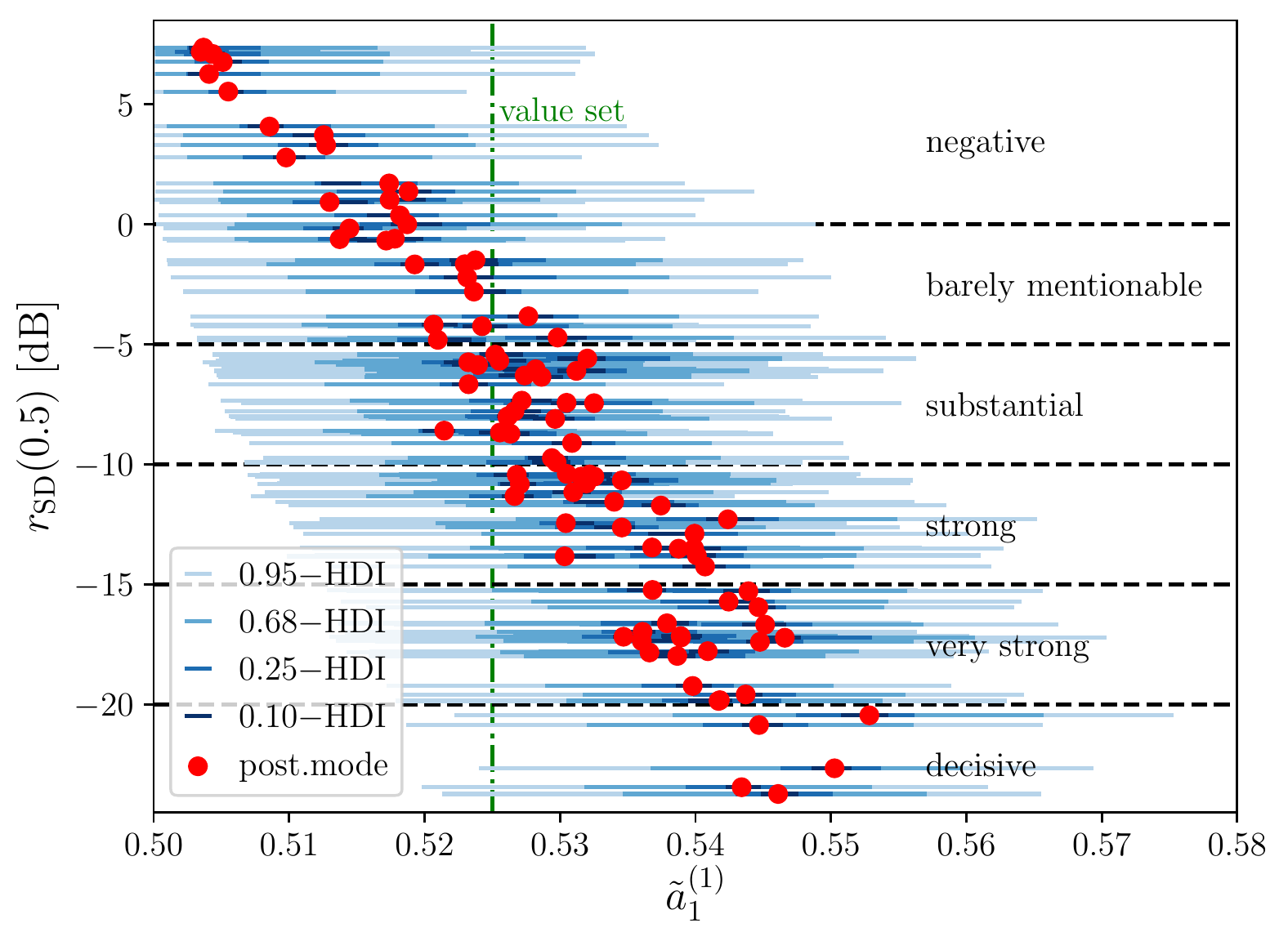}
		\caption{}\label{sfig_a1vsSD}
	\end{subfigure}
	\begin{subfigure}[b]{0.495\textwidth}
		\includegraphics[width=\textwidth]{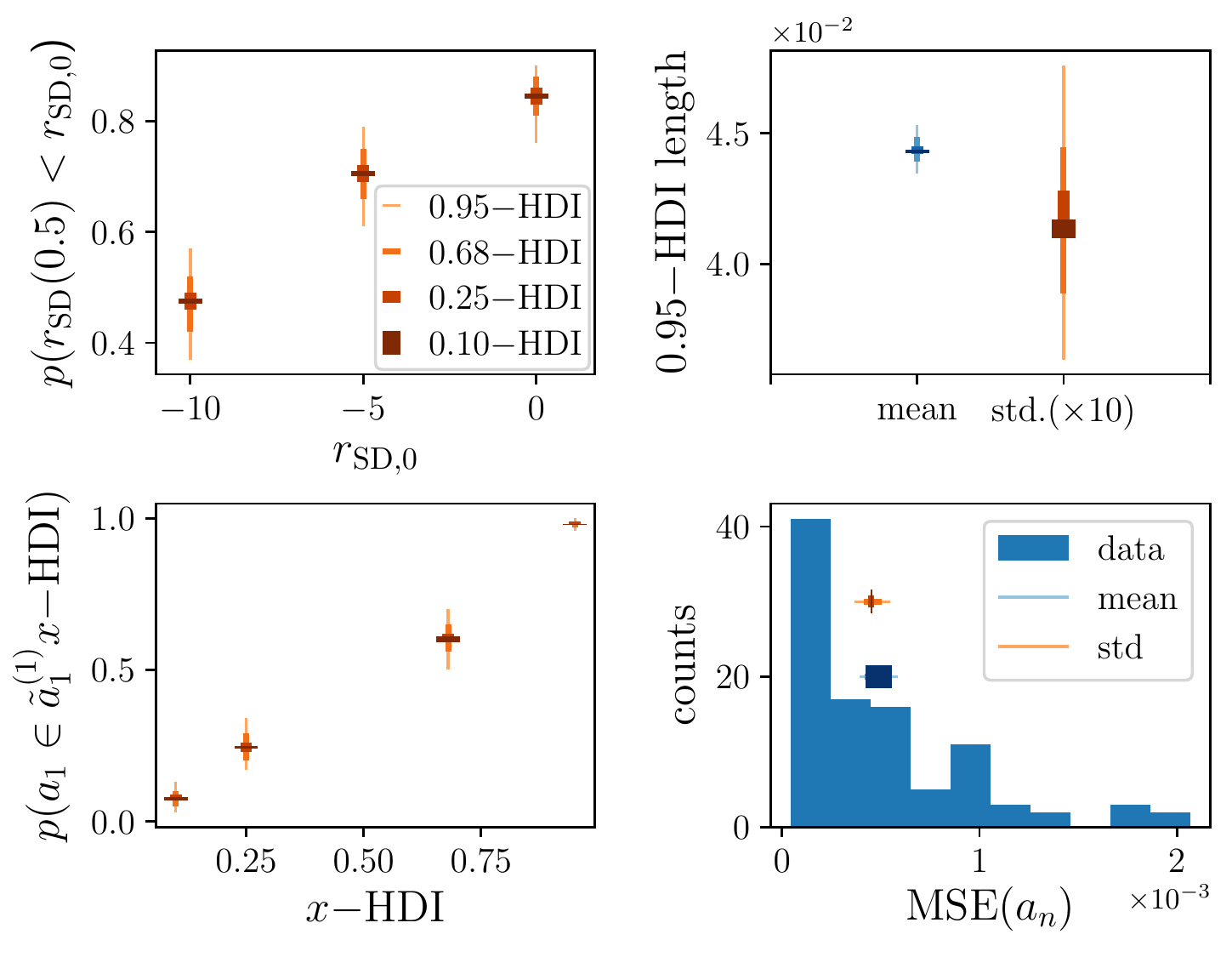}
		\caption{}\label{sfig_a1FigsMerit}
	\end{subfigure}
	\caption{Illustration of the figures of merit related to $a_1$ and the mean squared error of $\text{mode}(\tilde{a}_n^{(1)})$ with respect to $a_n$ (see text). (a) Modes and indicated HDIs of $\tilde{a}_1^{(1)}$ versus Savage-Dickey ratio for $100$ repetitions of the simulation and reconstruction with the same parameters as discussed in section \ref{secSingleSim}. The model $a_1$ is indicated with a green line and in black we indicate the common Savage-Dickey ratio interpretations \cite{Jeffreys1961}. (b) Results for the figures of merit. The top left, top right and bottom left panel show the result for $p(r_{\text{SD}}(0.5)< r_{\text{SD},0})$, the mean and standard deviation of the $0.95$-HDI length and $p(a_1\in \tilde{a}_1^{(1)} x\text{-HDI})$ respectively. The legend in the top left panel holds for all panels. The bottom left panel shows a histogram of the $100$ MSE$(a_n)$-values obtained from the reconstructions. The blue horizontal lines at $\text{counts}=20$ corresponds to the estimated mean of the distribution, and the orange lines at $\text{counts}=30$ represent the estimated standard deviation.}\label{figa1_FigsMerit100}
\end{figure}

In figure \ref{figa1_FigsMerit100} we show the figures of merit determined from these simulations. Figure \ref{sfig_a1vsSD} shows the mode and indicated HDIs for all $100$ reconstructions, plotted against $r_{\text{SD}}(0.5)$ as obtained from the reconstructions. As one would expect, the higher $\tilde{a}_1^{(1)}$, the lower $r_{\text{SD}}(0.5)$. From this figure we can easily determine $p(r_{\text{SD}}(0.5)< r_{\text{SD},0})$ from the fraction of reconstruction results below $r_{\text{SD},0}$. Also the mean length of the $0.95$-HDI and the probability that the $x$-HDI contains $a_1$ (which is indicated by the green line) are determined straightforwardly.\\
Estimations for these values are presented in figure \ref{sfig_a1FigsMerit}. For $p(r_{\text{SD}}(0.5)\leq r_{\text{SD},0})$ and $p(a_1\in \tilde{a}_1^{(1)} x\text{-HDI})$ the HDIs follow from considering the $100$ simulations as trials of a binomial experiment. Setting a uniform $\text{beta}(1,1)$-prior for both these probabilities, the posterior is a $\text{beta}(k+1,100+1)$-distribution, where $k$ is the amount of successful simulations (in which $r_{\text{SD}}(0.5)< r_{\text{SD},0})$ and $a_1\in \tilde{a}_1^{(1)} x\text{-HDI}$ respectively). The uncertainties for the other figures of merit follow from bootstrapping the simulation results $10^4$ times. It should be noted that for determining the mean $0.95$-HDI length of $\tilde{a}_1^{(1)}$ we only took into account the HDIs for which the minimum of the interval is larger than $0.501$. This excludes the reconstructions for which $\tilde{a}_1^{(1)}$ is close to $0.5$ (see figure \ref{sfig_a1vsSD}) resulting in a HDI of smaller size due to the $0.5$-bound.\\

\subsection{Optimum $\vec{\bar{N}}$, $D$ and $T$}\label{ssecOptNDT}
\begin{figure}[tbp]
	\centering
	\begin{subfigure}[b]{0.495\textwidth}
		\includegraphics[width=\textwidth]{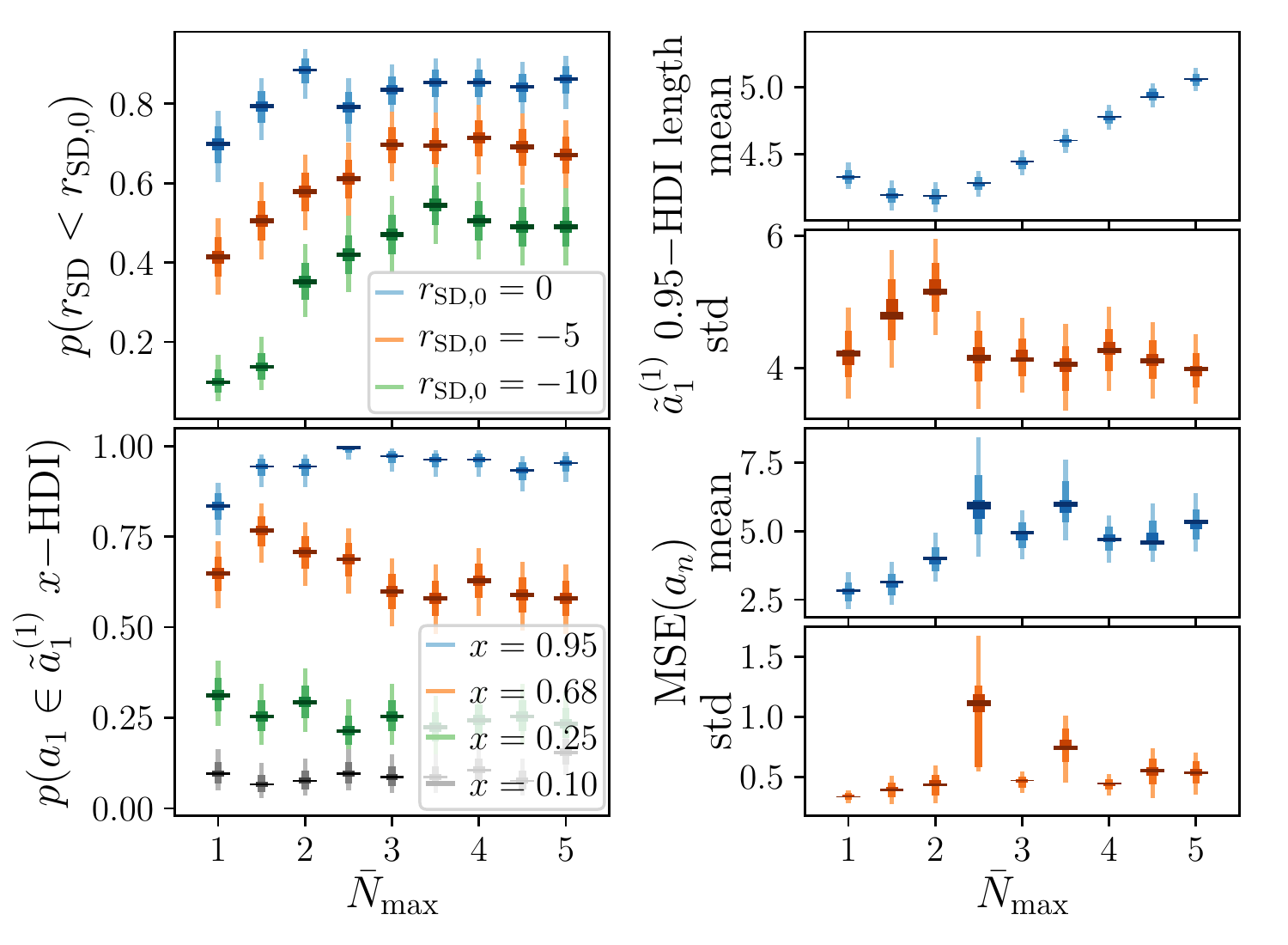}
		\caption{}\label{sfigMerit_Nmin10}
	\end{subfigure}
	\begin{subfigure}[b]{0.495\textwidth}
		\includegraphics[width=\textwidth]{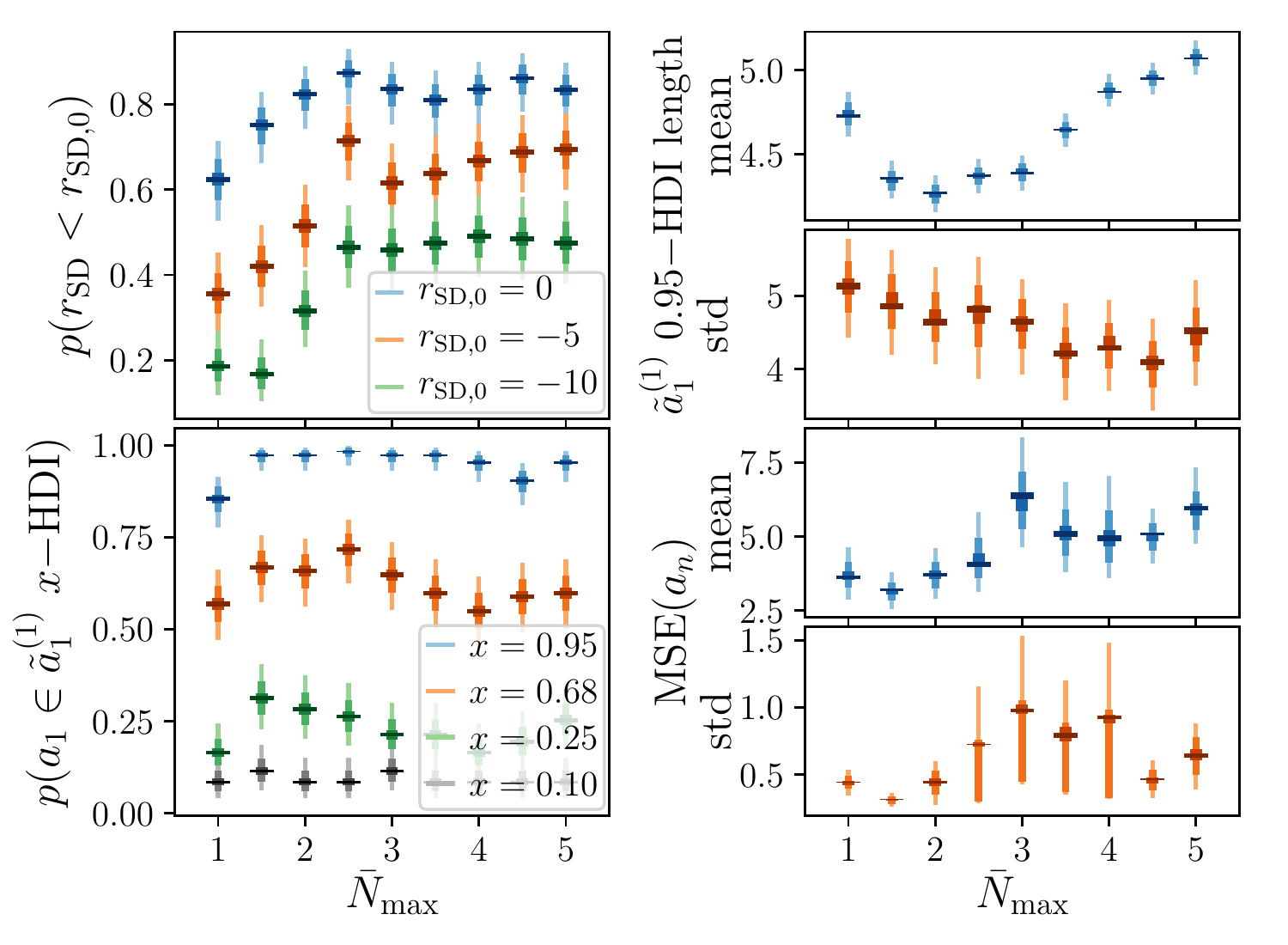}
		\caption{}\label{sfigMerit_Nmin05}
	\end{subfigure}
	\caption{Figures of merit estimated from $100$ simulations and reconstructions with (a) $\bar{N}_{\text{min}}=1.0$ and (b) $\bar{N}_{\text{min}}=0.5$. $p_1=0.05$, $D=5$ and $T=1000$, while we vary $\bar{N}_{\text{max}}$. The top left panel shows the results for $p(r_{\text{SD}}(0.5)\leq r_{\text{SD},0})$. It this probability initially increases linearly up to a critical $\bar{N}_{\text{max}}$ (Here $\bar{N}_{\text{max}}\approx3$) after which it remains constant. The top right panel shows the mean and standard deviation of the $0.95$-HDI length of $\tilde{a}_1^{(1)}$. The HDI length shows a clear minimum around $\bar{N}_{\text{max}}\approx2$. In the panel bottom left one observes $p(a_1\in \tilde{a}_1^{(1)} x\text{-HDI})$. There might be a slight preference for $\bar{N}_{\text{max}}\approx 1.5$, around which the probability is highest. In the bottom right panel we plot $\text{MSE}(a_n)$. The mean square error is expected to be lowest for lower values of $\bar{N}_{\text{max}}$. For both subfigures we observe the same behaviour, implying the chosen $\bar{N}_{\text{min}}$ is not critical.}\label{figMerit_Nmax}
\end{figure}

Using the figures of merit defined in previous section, we perform simulations and reconstructions to determine the optimum $\vec{\bar{N}}$, $D$ and $T$. For each experimental setting $(\vec{\bar{N}},D,T)$, we perform $100$ simulations and reconstructions to estimate the figures of merit.\\
In figure \ref{figMerit_Nmax} one can observe the results for setting $\bar{N}_{\text{min}}=1.0$ and $0.5$ respectively, while varying $\bar{N}_{\text{max}}$. Here $p_1=0.05$, $D=5$ and $T=1000$ as before. As can be seen, the behaviour of the four figures of merit does hardly seem to depend on $\bar{N}_{\text{min}}$. Interestingly, $p(r_{\text{SD}}(0.5)<r_{\text{SD},0})$ rises linearly for low $\bar{N}_{\text{max}}$ after which it hits a plateau at $\bar{N}_{\text{max}}\approx 3$. This critical value does not correspond to the optimum values indicated by the other figures of merit. This is most clear for the mean $0.95$-HDI length, which is minimised for $\bar{N}_{\text{max}}\approx 2$. Apart from that, $p(a_1\in \tilde{a}_1^{(1)} x\text{-HDI})$ also seems to favour a lower value for $\bar{N}_{\text{max}}$, $\bar{N}_{\text{max}}\approx 1.5$, although only weakly. $\text{MSE}(a_n)$ is lowest for the minimum values of $\bar{N}_{\text{max}}$ considered.\\
It becomes clear that the reconstruction results are influenced by two processes: first, if $\bar{N}_{\text{max}}$ is only slightly higher than $\bar{N}_{\text{min}}$, the range of photon numbers sent during the experiment is small. This implies there are relatively many trials per photon number $n$, such that the corresponding $\tilde{a}_n^{(1)}$s are determined from many trials. On the other hand, if $\bar{N}_{\text{max}}$ is much higher than $\bar{N}_{\text{min}}$, there are more $\tilde{a}_n$ relevant in the experiment, and therefore less trials per $n$ available. However, because there are more photon numbers relevant, the function describing $a_n$ can be better determined. It is the balance between these processes that determines the results.\\

\begin{figure}[tbhp]
	\centering
	\begin{minipage}[t]{0.50\textwidth}
		\includegraphics[width=0.99\textwidth]{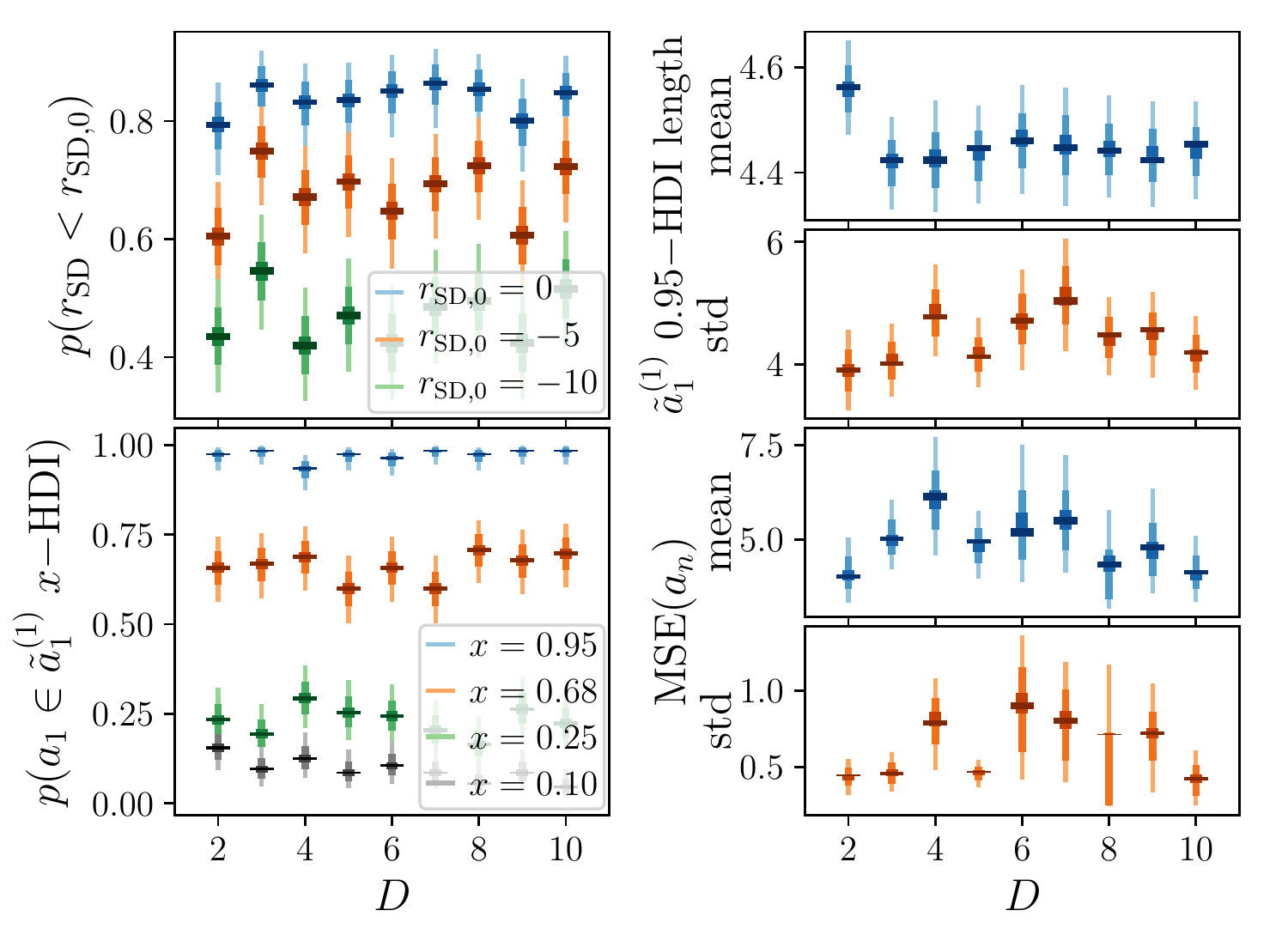}
		\caption{Figures of merit estimated from $100$ simulations and reconstructions with $\bar{N}_{\text{min}}=1$ and $\bar{N}_{\text{max}}=3$, while varying $D$ and keeping the total number of trials constant at $D\cdot T=5000$. $p_1=0.05$. It does not seem to matter how the (equidistant) data points $D$ are chosen in the range $[N_{\text{min}},N_{\text{max}}]$. Only the $0.95$-HDI length favours $D\geq 3$.} \label{figMerit_DT}
	\end{minipage}\hfill
	\begin{minipage}[t]{0.49\textwidth}
		\includegraphics[width=0.99\textwidth]{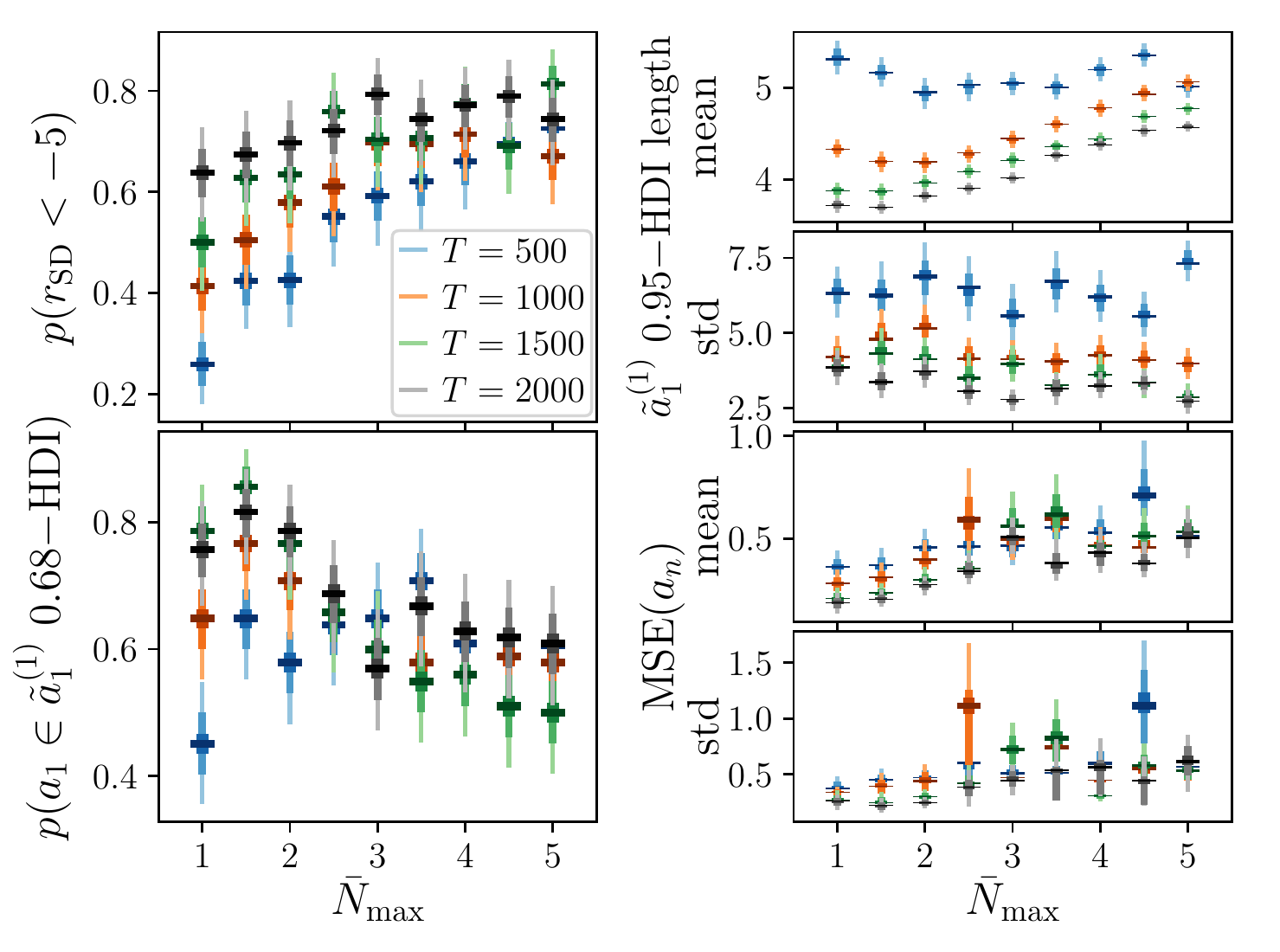}
		\caption{Figures of merit estimated from $100$ simulations and reconstructions with $\bar{N}_{\text{min}}=1$, $D=5$ and $p _1=0.05$. We vary the number of trials per data point $T$. $p(r_{\text{SD}}(0.5)<-5)$ in the top left panel shows the same behaviour as before (approximately linear rise to maximum value, then constant). Although the constant maximum $p(r_{\text{SD}}(0.5)<-5)$ seems to rise only slightly with $T$, the slope of the linear rise decreases with $T$. In the top right panel is is observed that the $0.95$-HDI length of $\tilde{a}_1^{(1)}$ decreases with increasing $T$ as one would expect. The minimum HDI-length is found at decreasing $\bar{N}_{\text{max}}$ with increasing $T$. For $p(a_1\in \tilde{a}_1^{(1)} x\text{-HDI})$, plotted in the lower left panel, we observe the same as for $p(r_{\text{SD}}(0.5)<-5)$: for low values of $\bar{N}_{\text{max}}$ the figure of merit increases with $T$, whereas for higher $\bar{N}_{\text{max}}$ the results seem to remain constant with $T$. In $\text{MSE}(a_n)$, bottom right, it is observed that this figure of merit improves with $T$, although only slightly.}\label{figMerit_Ttot}
	\end{minipage}
\end{figure}

So far we have considered $D=5$ and $T=1000$ (i.e. $5000$ trials in total). Let us now fix the amount of trials to $5000$ and vary $D$ (thus $T$). As can be observed in figure \ref{figMerit_DT}, it does not seem to matter much whether one chooses few data point with many trials per data point, or rather many data points with relatively few trials per data point. Only the mean $0.95$-HDI length favours $D\geq 3$. This result can be understood since the amount of trials per photon number will be more or less constant for every combination of $D$ and $T$ if $D\cdot T$ is kept constant.\\

Next, we consider the amount of trials. In figure \ref{figMerit_Ttot} we present our results on the figures of merit while varying $T$. For all $T$, we observe the same behaviour in $p(r_{\text{SD}}(0.5)<-5)$ as before. However, upon increasing $T$ we observe that the slope of the linear part of the curves decreases, while the critical $\bar{N}_{\text{max}}$ at which $p(r_{\text{SD}}(0.5)<-5)$ becomes constant remains the same, approximately $3.0$. Interestingly, it seems this constant value to which $p(r_{\text{SD}}(0.5)<-5)$ converges does not depend heavily on $T$, although it seems to rise slightly. We expect that if more simulations (and reconstructions) would be performed, such a dependence would become clear. The same holds for $p(a_1\in \tilde{a}_1^{(1)} x\text{-HDI})$.\\ 
For the $0.95$-HDI length of $\tilde{a}_1^{(1)}$, we observe that this length decreases with $T$ as one would expect. The minimum length, however, shifts to lower $\bar{N}_{\text{max}}$ with increasing $T$. It should be noted that the range of the $0.95$-HDI length does not vary much over the whole range of $\bar{N}_{\text{max}}$. This also holds for $\text{MSE}(a_n)$.\\

Summarising, from our simulations we find the the optimal experimental parameters depend on the figure of merit one wants to optimise. This is a valuable observation for experimental design.
 
\subsection{Influence of model $p_1$}\label{ssecModelp1}
\begin{figure}[tbhp]
\begin{minipage}[t]{0.49\textwidth}
	\centering
		\includegraphics[width=\textwidth]{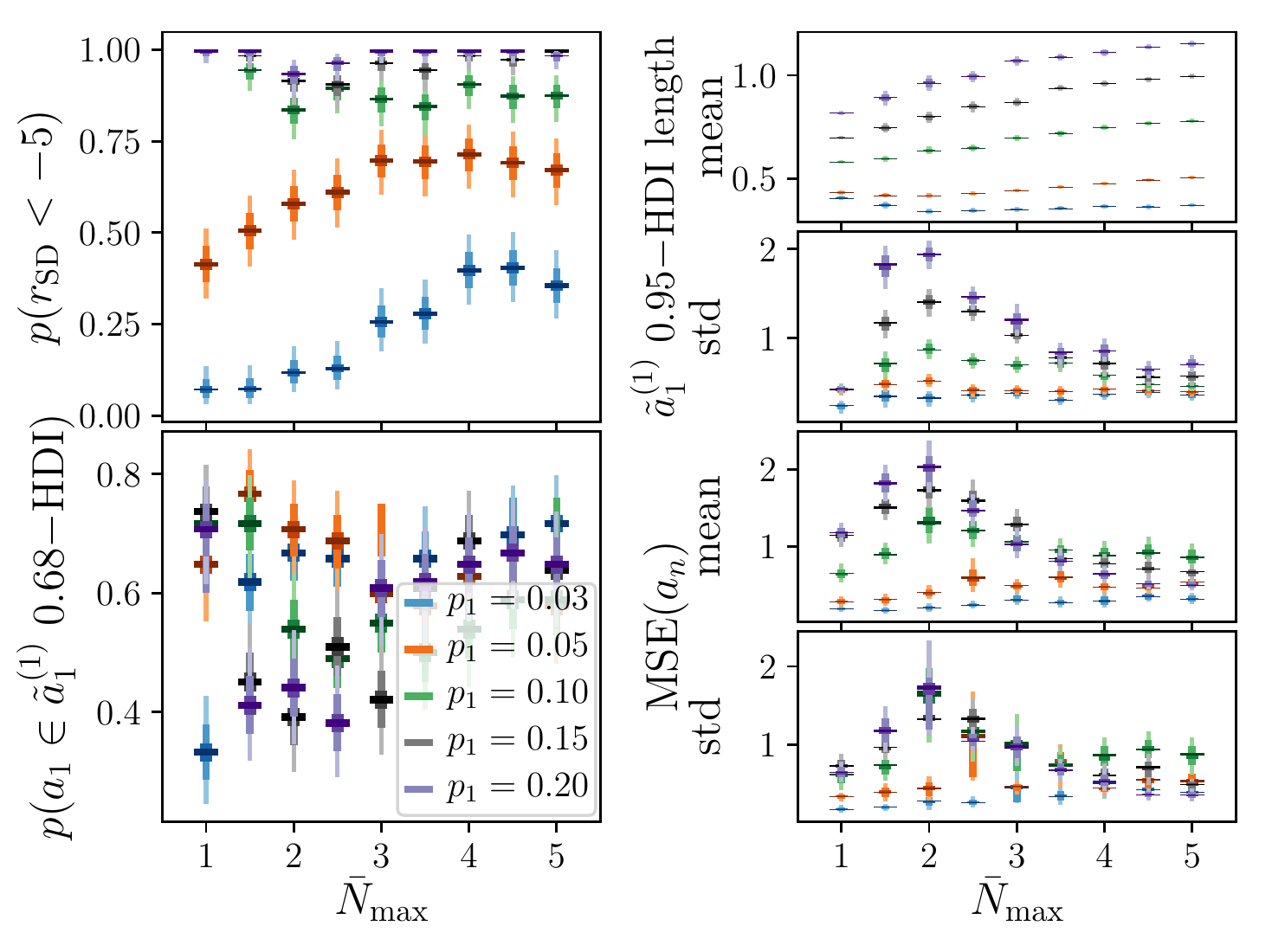}
		\caption{Figures of merit estimated from $100$ simulations and reconstructions with $\bar{N}_{\text{min}}=1$. $D=5$ and $T=1000$, while we vary the model value $p_1$. $p_1=0.03$ and $p_1=0.20$ correspond to the accuracies reported by \cite{Tinsleyetal2016} averaged over all trials and high-confidence trials respectively. In the top left panel it is observed for lower values of $p_1$ that $p(r_{\text{SD}}(0.5)<-5)$ increases up to a critical $\bar{N}_{\text{max}}$ after which the figure of merit remains constant. The critical $\bar{N}_{\text{max}}$ decreases with $p_1$. For the higher values of $p_1$, $p(r_{\text{SD}}(0.5)<-5)$ is approximately constant and close to $1$. As can be seen in the top right panel, the $0.95$-HDI length of $\tilde{a}_1^{(1)}$ increases with $p_1$, and the minimum value shifts to lower $\bar{N}_{\text{max}}$. $p(a_1\in \tilde{a}_1^{(1)} x\text{-HDI})$ in the bottom left panel shows no distinctive behaviour with $p_1$, whereas it becomes clear in the bottom right panel that $\text{MSE}(a_n)$ increases in general with $p_1$, and shows a distinct increase between $\bar{N}_{\text{max}}=1.0$ and $3.5$ for higher values of $p_1$. Based on these results, $\bar{N}_{\text{max}}=4.0$ is optimal.}\label{figMerit_p1}
\end{minipage}
\begin{minipage}[t]{0.49\textwidth}
	\centering
		\includegraphics[width=\textwidth]{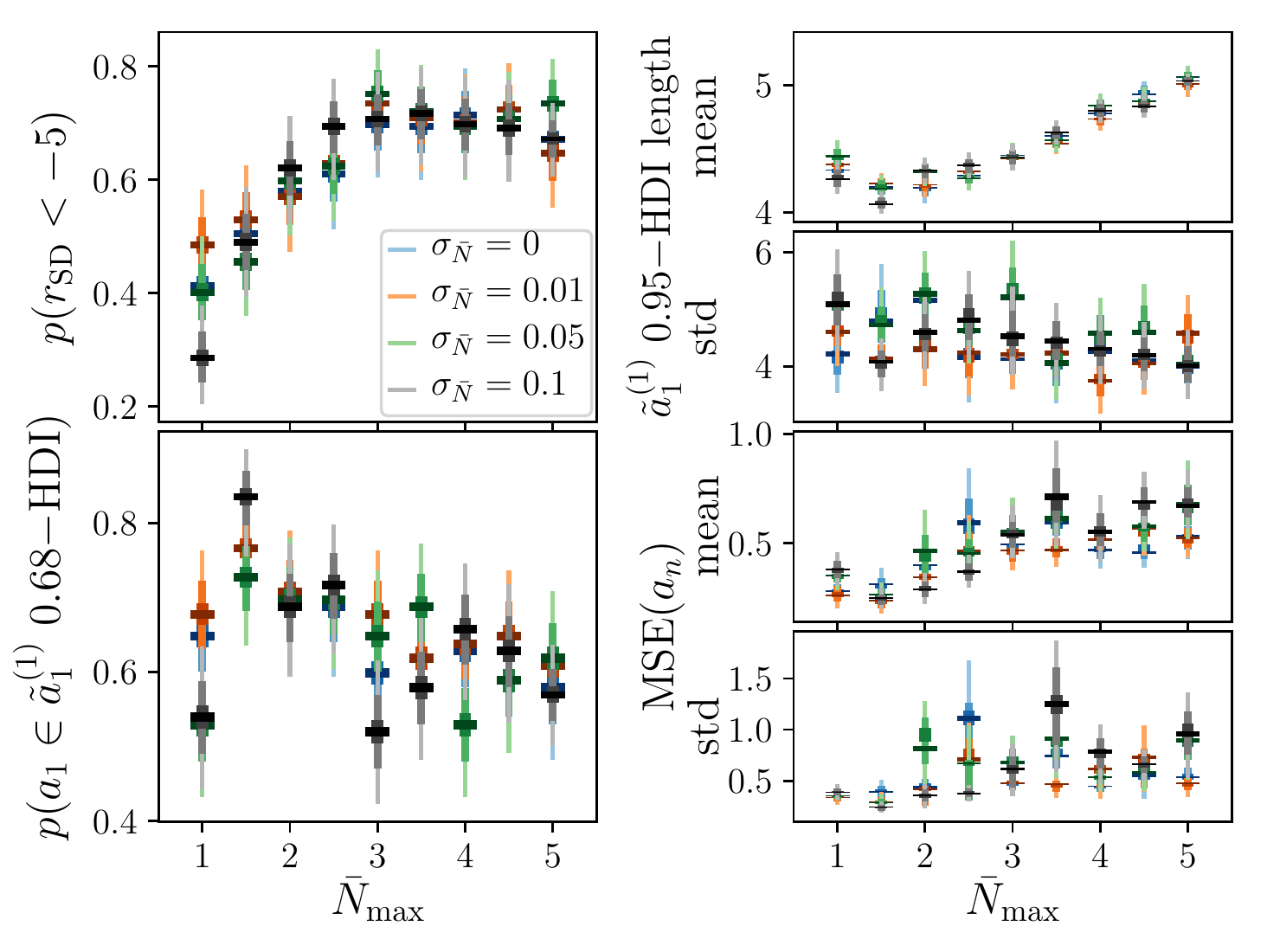}
		\caption{Figures of merit estimated from $100$ simulations and reconstructions with $\bar{N}_{\text{min}}=1$. $p_1=0.05$, $D=5$ and $T=1000$, while we vary the light source's noise: $\bar{N}_d=\bar{N}_{d,0}+\mathcal{N}(0,\sigma_{\bar{N}})$. For all of the figures of merit, it is observed that the amount of noise (up to $\sigma_{\bar{N}}=0.1$) does hardly influence the results.}\label{figMerit_dN}
\end{minipage}
\end{figure}

The visual perception model we have presented in section \ref{secEyeModel} depends on the single parameter $p_1$. We therefore study the influence this parameter has on our figures of merit. In figure \ref{figMerit_p1}, we present the results for $p_1=0.03$ up to $p_1=0.20$, which correspond to the results obtained in \cite{Tinsleyetal2016} for respectively all trials and high-confidence trials.\\
As can be seen in the figure, the figures of merit highly depend on $p_1$. $p(r_{\text{SD}}(0.5)<-5)$ is seen to show the same behaviour as before, although $\bar{N}_{\text{max}}$ at which $p(r_{\text{SD}}(0.5)<-5)$ becomes constant is observed to decrease with increasing $p_1$. For $p_1=0.10$ and higher, the linear part has disappeared and $p(r_{\text{SD}}(0.5)<-5)\approx 1$ for the whole range of $\bar{N}_{\text{max}}$.\\
The $0.95$-HDI length of $\tilde{a}_1^{(1)}$ and $\text{MSE}(a_n)$ are seen to increase generally with increasing $p_1$. This can be understood as the difference  between $a_{n}$ and $a_{n+1}$ increases for increasing $p_1$. This implies that in the reconstructions $\tilde{a}_{n+1}^{(1)}$ ``confines'' $\tilde{a}_n^{(1)}$ to a smaller extent, such that $\tilde{a}_n^{(1)}$ is more free to explore the parameter space.

\subsection{Influence of noise on $\vec{\bar{N}}$}\label{ssecLightNoise}
Finally, we consider the influence of the amount of noise on $\vec{\bar{N}}$. 
We set $\sigma_{\bar{N},d}=\sigma_{\bar{N}}$ and vary $\sigma_{\bar{N}}$ from $0$ to $0.1$. As can be observed in figure \ref{figMerit_dN}, the figures of merit are hardly influenced by the amount of noise added, as discussed in the main text.
\end{document}